# Improved Approximation Lower Bounds for Vertex Cover on Power Law Graphs and Some Generalizations

Mikael Gast[*]   Mathias Hauptmann[†]   Marek Karpinski[‡]

**Abstract**

We prove new explicit inapproximability results for the Vertex Cover Problem on the Power Law Graphs and some functional generalizations of that class of graphs. Our results depend on special bounded degree amplifier constructions for those classes of graphs and could be also of independent interest.

## 1  Introduction

Recently the study of *large-scale* real-world networks revealed common topological signatures and statistical features that are not easily captured by classical *uniform* random graphs—such as generated by the $G(n,p)$-model due to Erdős and Rényi [ER60]. As of 1999 Kumar et al. [Kum+00; Bro+00], Kleinberg et al. [Kle+99; KL01] and Faloutsos, Faloutsos, and Faloutsos [FFF99; Sig+03] measured the degree sequence of the World-Wide Web and independently observed that it is well approximated by a *power-law distribution*, i.e. the number of nodes $y_i$ of a given degree $i$ is proportional to $i^{-\beta}$ where $\beta > 0$. This was later verified for a large number of existing real-world networks such as protein-protein interactions, gene regulatory networks, peer-to-peer networks, mobile call networks and social networks [JAB01; Gue+02; Ses+08; Eub+04]. In fact, power-law distributions had also been observed considerably earlier for the distribution of income, city sizes, word frequencies and for citations of academic (chemist) literature [Par96; Aue13; Est16; Lot26]. Besides these and other early investigations, the idea of associating power-law distributions with real-life systems (and its popularization) is generally attributed to Zipf [Zip50] and is therefore often referred to as an associated *Zipfian distribution* (also *heavy-tail distribution* or *Pareto distribution*).

The model of *preferential attachment* is most often referred to as the mechanism underlying the construction of the above graphs, featuring the role of evolutionary growth or rewiring processes. First mentioned and described by Yule in [Yul25], Simon in [Sim55] and de Solla Price in [Pri65; Pri76] this concept was introduced to a broader audience by Barabási and Albert in [BA99] and later was more rigorously and mathematically defined by Bollobás and Riordan in [BR05]. In this model, a newly introduced vertex will connect to already existing vertices with a probability depending on their current degree. This principle of network growth is therefore (customary) described as *"the rich get richer"* or *"preferential attachment"*. But preferential attachment is only one of several mechanisms that can produce graphs with power-law degree distributions, so called *Scale-Free Networks* or, especially, *Power-Law Graphs* (PLG).

---
[*]Dept. of Computer Science, University of Bonn and B-IT Research School. e-mail: `gast@cs.uni-bonn.de`
[†]Dept. of Computer Science, University of Bonn. e-mail: `hauptman@cs.uni-bonn.de`
[‡]Dept. of Computer Science and the Hausdorff Center for Mathematics, University of Bonn. Research partially supported by the grant EXC59-1. e-mail: `marek@cs.uni-bonn.de`

Motivated by the behavior of massive graphs derived from data in telecommunications, Aiello, Chung, and Lu [ACL00; ACL01] proposed a model that ensures a power-law degree distribution by fixing a degree sequence via two parameters $\alpha, \beta$ and then to take the space of random graphs with this degree sequence. Thus their approach somehow complements the above generation models in that it does not aim to explain how power-laws arise, but, given that a graph has a power-law degree sequence, allows to derive structural properties and statistical features which hold with asymptotically high probability (probability tending to 1 while the size of the graph approaches $\infty$). Furthermore the derived results are true not only for certain instances of the random graph model, but for the majority of graphs with the given degree sequence. This model will be referred to as the $(\alpha, \beta)$-*model* or *ACL-model* for random PLG. The corresponding graph class will be called $(\alpha, \beta)$-PLG.

Apart from these modeling approaches there exists practical evidence that combinatorial optimization in PLG is easier than in general graphs [PL01; GMS03; Eub+04; KSG06]. In [GH12] Gast and Hauptmann construct an approximation algorithm for the Minimum Vertex Cover Problem (Min-VC) with an expected approximation ratio of $2 - f(\beta)$ for random PLG in the ACL-model, where $f(\beta)$ is a strictly positive function of the model parameter $\beta$. Note that $f(\beta)$ does not depend on the size $|V|$ of the graph and thus – for large graph sizes – falls below current upper bounds for Min-VC in general graphs (which is $2 - \Theta\left(\frac{1}{\sqrt{\log n}}\right)$ as stated by Karakostas in [Kar09]).

Contrasting this Ferrante, Pandurangan, and Park [FPP08] and Shen, Nguyen, and Thai [SNT10; She+12] studied the approximation hardness of certain optimization problems in combinatorial PLG in the ACL-model and showed NP-hardness and APX-hardness for classical problems such as Minimum Vertex Cover (Min-VC), Maximum Independent Set (Max-IS) and Minimum Dominating Set (Min-DS).

Let us give a brief summary of previous results regarding approximation lower bounds of Min-VC in $(\alpha, \beta)$-PLG. In [FPP08] Ferrante, Pandurangan, and Park showed NP-hardness of Max-IS, Min-DS, and Min-VC in simple disconnected $(\alpha, \beta)$-PLG for $\beta > 0$. In [SNT10] Shen, Nguyen, and Thai proved APX-hardness of the same problems in disconnected $(\alpha, \beta)$-PLG *multigraphs* for $\beta > 1$. Table 1 summarizes the results of Shen et al., showing the inapproximability factors of Max-IS, Min-DS, and Min-VC in simple disconnected $(\alpha, \beta)$-PLG and disconnected $(\alpha, \beta)$-PLG multigraphs presented in [She+12] for $\beta > 1$. Note that in particular, the above results do not directly imply hardness and inapproximability in *connected* PLG and for power-law exponents $\beta \leqslant 1$. An open question regarding the hardness of Max-IS, Min-DS, and Min-VC in *connected* $(\alpha, \beta)$-PLG is posed in [FPP08].

| Problem | $(\alpha, \beta)$-PLG multigraphs | $(\alpha, \beta)$-PLG |
|---|---|---|
| Max-IS | $1 + \frac{1}{140(2\zeta(\beta)3^\beta - 1)} - \varepsilon$ | $1 + \frac{1}{1120\zeta(\beta)3^\beta} - \varepsilon$ |
| Min-DS | $1 + \frac{1}{390(2\zeta(\beta)3^\beta - 1)}$ | $1 + \frac{1}{3120\zeta(\beta)3^\beta}$ |
| Min-VC | $1 + \frac{2\left(1 - (2 + o_c(1))\frac{\log \log c}{\log c}\right)}{\left(\zeta(\beta)c^\beta + c^{\frac{1}{\beta}}\right)(c-1)}$ | $1 + \frac{2 - (2 + o_c(1))\frac{\log \log c}{\log c}}{2\zeta(\beta)c^\beta(c+1)}$ |

**Table 1:** Inapproximability factors of Max-IS and Min-DS under condition $\mathsf{P} \neq \mathsf{NP}$, Min-VC under UGC in disconnected power-law graphs with $\beta > 1$ due to [She+12].

In this paper we show the APX-hardness of Min-VC in *connected* $(\alpha, \beta)$-PLG multigraphs for $0 < \beta < \beta_{\max} \approx 2.48$ and give explicit approximation lower bounds for this problem. For $\beta > \beta_{\max}$, $(\alpha, \beta)$-PLG are not connected anymore. Our results are based on the construction of special bounded degree amplifiers. A similar method has already been used in [BK99; BK01;

BK03] to obtain explicit lower bounds for the approximability of bounded degree and small occurrence optimization problems. Our reductions consist of multigraph embeddings of bounded degree graphs into $(\alpha, \beta)$-PLG, based on appropriate wheel constructions. We also extend the model of $(\alpha, \beta)$-PLG and consider degree distributions where $\beta$ is of the form $\beta = 1 \pm 1/f(n)$ for a sufficiently fast growing function $f(n)$. These distributions converge to those of $(\alpha, \beta)$-PLG for $\beta = 1$ and can be seen as a combinatorial variant of the preferential attachment PLG.

*The paper is organized as follows:* In Section 2 we give the formal definition of $(\alpha, \beta)$-Power-Law Graphs in the ACL-model. Section 3 gives an outline of the methodology of the reduction, i.e. the algorithm for the construction of the wheel and the general embedding technique for $d$-bounded graphs into connected $(\alpha, \beta)$-PLG. In Section 4 we give the detailed description of our reduction from MIN-VC in $d$-bounded degree graphs to MIN-VC in connected $(\alpha, \beta)$-PLG for the parameter $\beta$ in the interval $\beta \in (1, \beta_{\max}]$. Section 5 deals with the case $\beta \in (0, 1]$ and gives the details of the reduction of MIN-VC in $d$-bounded degree graphs to MIN-VC in $d$-bounded degree graphs which provide a *perfect matching* and then to MIN-VC in connected $(\alpha, \beta)$-PLG. Furthermore, we give a thorough error term analysis for this case. Figure 1 shows the global organization of the paper, pointing to the different ranges and the phase transitions of the parameter $\beta$.

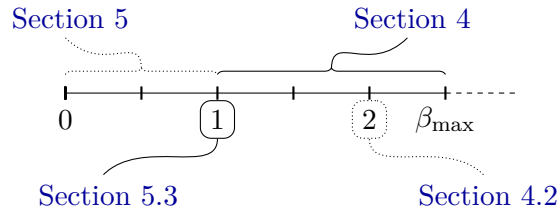

**Figure 1:** Organization of the paper with respect to the phase transitions and different ranges of the model parameter $\beta$.

Furthermore, we prove explicit lower bounds for the approximability of MIN-VC in $(\alpha, \beta)$-PLG which only depend on the degree bound $d$, the parameter $\beta$ and on the lower bounds $\varepsilon_d$ for $d$-bounded MIN-VC. The resulting inapproximability factors are summarized in Table 2.

|  | $\beta \in (0, 1]$ | $\beta \in (1, 2]$ | $\beta \in (2, \beta_{\max}]$ |
|---|---|---|---|
| Inapprox. factors | $1 + \frac{\varepsilon_d}{1+2d}$ | $1 + \frac{\varepsilon_d}{1+d \cdot \frac{(\zeta(\beta)-1) \cdot (d+1)^\beta - 1}{2}}$ | $1 + \frac{\varepsilon_d}{1+d(d+1)^\beta \left( \frac{1}{2^{\beta+1}} + \zeta(\beta) - 1 - \frac{1}{2^\beta} - \frac{1}{(d+1)^\beta} \right)}$ |

**Table 2:** Inapproximability factors of MIN-VC in connected $(\alpha, \beta)$-PLG multigraphs for the three half-open intervals between 0 and $\beta_{\max}$.

Figure 2 shows a comparison of the resulting inapproximability factors for various $d$ and corresponding lower bounds $(\varepsilon_d)$ for $d$-bounded MIN-VC on the subintervals $\beta \in (0, 1]$, $\beta \in (1, 2]$ and $\beta \in (2, \beta_{\max}]$. Figure 3 is a plot of the inapproximability factors over the whole interval $(0, \beta_{\max}]$ for $d = 3$ and $\varepsilon_d = 7/6$, where the points of discontinuity (*jumps*) correspond to the phase transitions at $\beta = 1$ and $\beta = 2$.

In Section 6 and Section 7 we consider an extension of the ACL-Model where the parameter $\beta$ is not constant but a function of the size of the graph. We give explicit approximation lower bounds for the case $\beta = 1 - \frac{1}{f(n)}$ and $\beta = 1 + \frac{1}{f(n)}$ where $f(n)$ is a sufficiently fast growing unbounded function. This extension is motivated by the fact that in the preferential attachment models, the degree distribution only converges to a power-law distribution in the limit. Hence this model can be seen as a combinatorial version of the preferential attachment PLGs which allows to derive approximation hardness results. It can be seen as a dynamic power-law model which allows the power-law exponent to vary over time as a function of recent data, i.e. the size of a growing network or time.

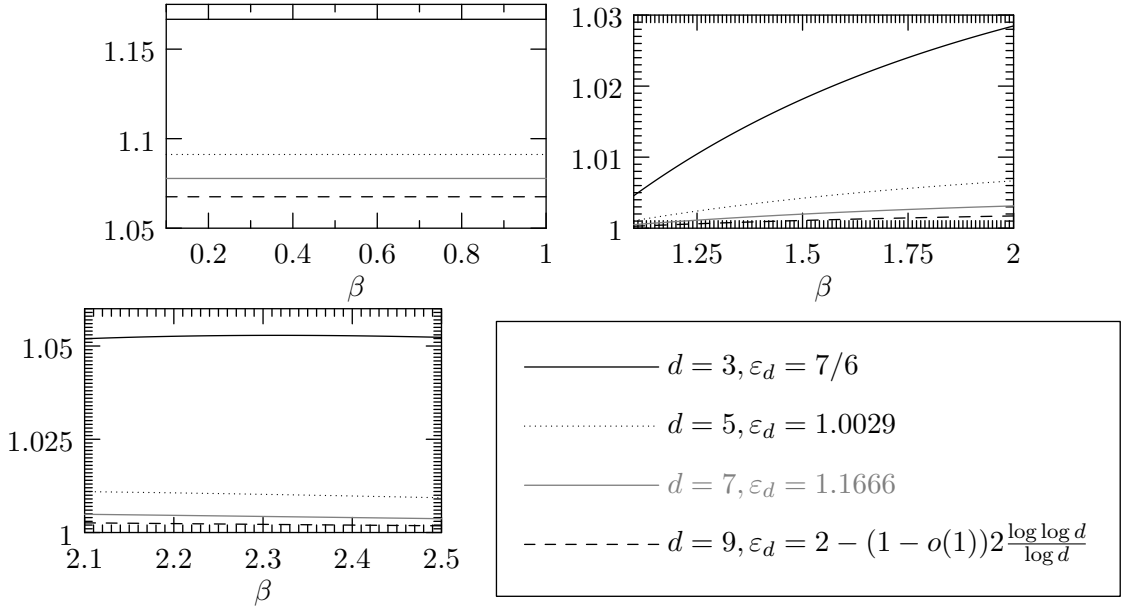

**Figure 2:** Comparison of the inapproximability factors for MIN-VC in connected $(\alpha, \beta)$-PLG multigraphs for various $d$ and $\varepsilon_d$ on the subintervals $\beta \in (0, 1]$, $\beta \in (1, 2]$ and $\beta \in (2, \beta_{\max})$.

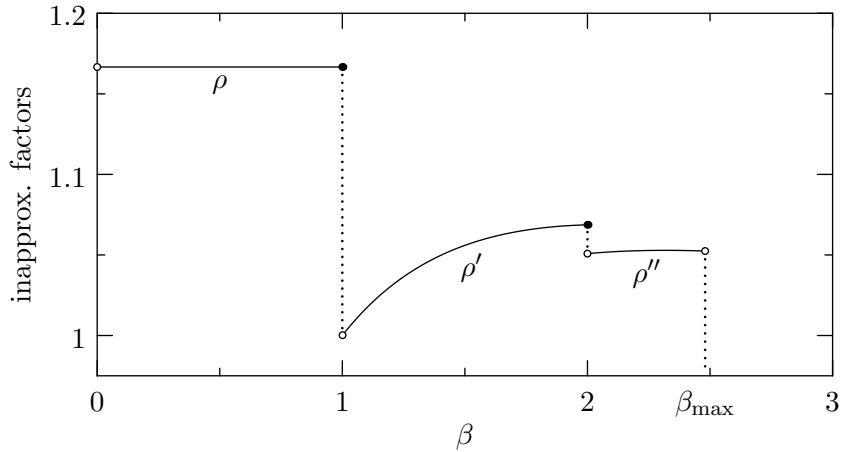

**Figure 3:** Plot of the inapproximability factor for MIN-VC in connected $(\alpha, \beta)$-PLG multigraphs for $d = 3$ and $\varepsilon_d = 7/6$ on the interval $(0, \beta_{\max})$.

## 2 $(\alpha, \beta)$-Power-Law Graphs

In this section we give the formal definition of $(\alpha, \beta)$-Power Law Graphs and describe the random PLG-model proposed by Aiello, Chung, and Lu [ACL01]. Furthermore we give a formula for the expected cut-size in this model in terms of the degree sequences of the two sides of the cut. This will give support to our constructions, which are basically embeddings of bounded degree graphs $G_d$ into $(\alpha, \beta)$-PLG $\mathcal{G}_{\alpha,\beta}$ such that the size of the cut between $G_d$ and $\mathcal{G}_{\alpha,\beta} \setminus G_d$ is at least linear in the size of $G_d$.

**Definition 1** (($\alpha, \beta$)-*Power Law Graphs* [ACL01]). An $(\alpha, \beta)$-PLG is an undirected multigraph (possibly containing self-loops) $\mathcal{G}_{\alpha,\beta} = (V, E)$ of maximum degree $\Delta = \lfloor e^{\alpha/\beta} \rfloor$ such that for $i = 1, \ldots, \Delta$, $\mathcal{G}_{\alpha,\beta}$ contains $y_i$ nodes of degree $i$, where

$$y_i = \begin{cases} \left\lfloor \frac{e^\alpha}{i^\beta} \right\rfloor & \text{if } i > 1 \text{ or } \sum_{i=1}^{\Delta} \left\lfloor \frac{e^\alpha}{i^\beta} \right\rfloor \text{ is even} \\ \lfloor e^\alpha \rfloor + 1 & \text{otherwise.} \end{cases}$$

Here, $i$ and $y_i$ satisfy $\log y_i = \alpha - \beta \log i$. Furthermore, $\alpha$ is the logarithm of the size of the graph and $\beta$ is the log-log growth rate. The number $n = \sum_{i=1}^{\lfloor e^{\alpha/\beta} \rfloor} \left\lfloor \frac{e^\alpha}{i^\beta} \right\rfloor$ of nodes and the number $m = \frac{1}{2} \sum_{i=1}^{\lfloor e^{\alpha/\beta} \rfloor} \left\lfloor \frac{e^\alpha}{i^\beta} \right\rfloor$ of edges of an $(\alpha, \beta)$-PLG satisfy

$$n \approx \begin{cases} \zeta(\beta) e^\alpha & \text{if } \beta > 1 \\ \alpha e^\alpha & \text{if } \beta = 1 \\ \frac{e^{\frac{\alpha}{\beta}}}{1-\beta} & \text{if } 0 < \beta < 1 \end{cases} \qquad m \approx \begin{cases} \frac{1}{2} \zeta(\beta - 1) e^\alpha & \text{if } \beta > 2 \\ \frac{1}{4} \alpha e^\alpha & \text{if } \beta = 2 \\ \frac{1}{2} \frac{e^{\frac{2\alpha}{\beta}}}{2-\beta} & \text{if } 0 < \beta < 2 \end{cases}$$

As already stated in [ACL01], the rounding error (which results from working with the real numbers $\frac{e^\alpha}{i^\beta}$, $e^{\frac{\alpha}{\beta}}$ instead of their integer counterparts) is a lower order term in the case $\beta > 2$. For our construction, the crucial point will be to give a precise estimate of the rounding errors in the case $\beta \leqslant 2$.

The *random graph model* for $(\alpha, \beta)$-PLG proposed by Aiello, Chung, and Lu [ACL01] is the distribution $P(\alpha, \beta)$ on the set of all $(\alpha, \beta)$-PLG which is obtained in the following way:

1. Generate a set $L$ of $\mathsf{deg}(v)$ distinct copies of each vertex $v \in V$.

2. Generate a random matching on the elements of $L$.

3. For each pair of vertices $u$ and $v$, the number of edges joining $u$ and $v$ in $G$ is equal to the number of edges in the matching of $L$ which join copies of $u$ to copies of $v$ (see Figure 4).

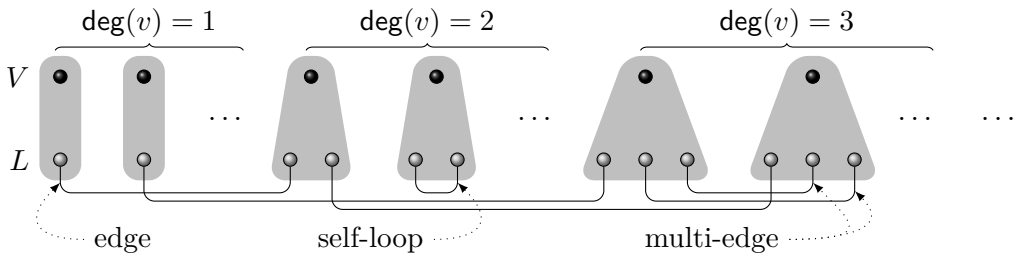

**Figure 4:** Generation of edges, self-loops and multi-edges in the random graph model for $(\alpha, \beta)$-PLG via random matching of vertex copies in $L$.

We will now deal with the expected cut sizes in the random PLG-model $P(\alpha, \beta)$. For a given degree sequence $d_1, \ldots, d_{n'}$, let the set $L$ be defined as $L = \bigcup_{i=1}^{n'} L_i$ with $L_i = \{v_{i,1}, \ldots, v_{i,d_i}\}$ Note that in the case of an $(\alpha, \beta)$-PLG, the degree sequence is

$$\underbrace{1 \ldots 1}_{\lfloor e^\alpha \rfloor} \ldots \underbrace{j \ldots j}_{\left\lfloor \frac{e^\alpha}{j^\beta} \right\rfloor} \ldots \Delta = \lfloor e^{\alpha/\beta} \rfloor.$$

Now let $n = \sum_{j=1}^{n'} d_j$, and let $\mathcal{M}^n$ denote the set of all matchings on the set $L$. Furthermore, for $v, v' \in L$, let $\mathcal{M}^n_{v,v'}$ be the set of all matchings which contain the edge $\{v, v'\}$. We have $|\mathcal{M}^n| = \frac{n!}{2^{n/2} \cdot (\frac{n}{2})!}$ and $|\mathcal{M}^n_{v,v'}| = \frac{(n-2)!}{2^{(n-2)/2} \cdot (\frac{n-2}{2})!}$.

Hence the probability for an edge $e = \{v, v'\}$ to be an element of a random matching $M$ over $n$ vertices is

$$\Pr(e = \{v_i, v_j\} \in \mathcal{M}^n) = \frac{|\mathcal{M}^n_{v,v'}|}{|\mathcal{M}^n|} = \frac{\frac{(n-2)!}{2^{(n-2)/2} \cdot (\frac{n-2}{2})!}}{\frac{n!}{2^{n/2} \cdot (\frac{n}{2})!}} = \frac{1}{n-1}.$$

Thus we obtain the following result.

**Lemma 1.** *Consider the random PLG-model $P(\alpha, \beta)$ and let $A, B$ be disjoint subsets of vertices of the resulting PLG. Then the expected number of edges between $A$ and $B$ is*

$$\mathbb{E}(\#edges \text{ between } A \text{ and } B) = \frac{\left(\sum_{w \in A} \deg(w)\right) \left(\sum_{u \in B} \deg(u)\right)}{\sum_{v \in V} \deg(v)}$$

## 3 Outline of the Method

For each parameter $\beta \in [0, \beta_{\max})$ as well as for the functional cases $\beta = 1 \pm \frac{1}{f(n)}$ we will construct a polynomial time reduction $R_\beta$ from MIN-VC$_d$ (MIN-VC in $d$-bounded graphs) to MIN-VC$_{\alpha,\beta}$. $R_\beta$ embeds any $d$-bounded graph $G_d$ into a multigraph $R_\beta(G_d) = \mathcal{G}_{\alpha,\beta}$ which consists of a multigraph copy $\mu(G_d)$ of $G_d$ attached to a multigraph wheel $W$. Furthermore all the degree 1 nodes of $\mathcal{G}_{\alpha,\beta}$ are attached to wheel nodes (amplifiers) of $W$. In any case the size of $\mathcal{G}_{\alpha,\beta}$ will be linear in the size of $G_d$. We let $\Gamma$ denote the set of neighbors of $G_d$ in the wheel $W$, and $W_1$ denotes the set of wheel nodes which are adjacent to at least one degree-1 node in $\mathcal{G}_{\alpha,\beta}$ (see Figure 5).

We will make use of the notion of an *interval* in a PLG. Let $\mathcal{G}_{\alpha,\beta} = (V, E)$ an $(\alpha, \beta)$-PLG. An *interval* of nodes in $\mathcal{G}_{\alpha,\beta}$ is a set $[a, b] = \{v \in V \mid a \leqslant \deg(v) \leqslant b\}$, where $1 \leqslant a \leqslant b \leqslant \Delta = \lfloor e^{\alpha/\beta} \rfloor$.

Due to the different behavior of the power-law distributions, we have to distinguish the two cases $1 < \beta < \beta_{\max} \approx 2.48$ and $0 < \beta \leqslant 1$.

For $1 < \beta < \beta_{\max}$ we construct the PLG $\mathcal{G}_{\alpha,\beta} = \mu(G_d) \cup W$ in such a way that the set $\Gamma$ of neighbors of $G_d$ in the wheel $W$ satisfies $\Gamma \subseteq W_1$ and $|\Gamma| = \Theta(n)$. This means every neighbor of nodes from $G_d$ in the wheel is also adjacent to at least one node of degree 1. Neighbors of degree-1 nodes have the property that every vertex cover either contains this node or all its degree-1 neighbors. This implies that any optimum vertex cover $C^{\text{OPT}}$ in $\mathcal{G}_{\alpha,\beta}$ contains the set $\Gamma$, and hence the intersection of $C^{\text{OPT}}$ with $\mu(G_d)$ corresponds to an optimum vertex cover $C_d^{\text{OPT}}$ in $G_d$.

In the case $0 < \beta \leqslant 1$ and also in the functional cases $\beta = 1 - \frac{1}{f(n)}$ and $\beta = 1 + \frac{1}{f(n)}$, the behavior of the power-law distributions is rather different. In these cases, the number of degree-1 nodes in $(\alpha, \beta)$-PLG is too small to attach a degree-1 node to every neighbor of $\mu(G_d)$ in $W$. Hence we cannot guarantee anymore that every optimum vertex cover in $\mathcal{G}_{\alpha,\beta}$ contains an optimum vertex cover in $G_d$. And another problem occurs: In order to obtain $|\mathcal{G}_{\alpha,\beta}| = O(|G_d|)$, the nodes of $\mu(G_d)$ must have high degree in $\mathcal{G}_{\alpha,\beta}$. Since the set $\Gamma$ is too small to realize this degree of nodes in $G_d$, we need to replace the edges of $G_d$ by sufficiently many multi-edges. In order to keep track of the node-degrees and to implement the power-law distribution, we will first map $G_d$ to a graph $\widetilde{G}_{d+2}$ which contains a perfect matching. This allows us to increase the node degrees inside $\widetilde{G}_{d+2}$ in a controlled manner, namely pairwise along the edges of a perfect matching. Then we construct $\mathcal{G}_{\alpha,\beta} = R_\beta(G_d)$ in such a way that $|\Gamma| = o(n)$. This means in the cases $0 < \beta \leqslant 1$ and in the functional cases $\beta = 1 - \frac{1}{f(n)}$ and $\beta = 1 + \frac{1}{f(n)}$, our reduction from MIN-VC$_d$ to MIN-VC$_{\alpha,\beta}$ is the composition of a reduction from MIN-VC$_d$ to MIN-VC$_{d+2}^{PM}$ – MIN-VC restricted to $(d+2)$-bounded degree graphs which provide a perfect matching – and a reduction from MIN-VC$_{d+2}^{PM}$ to MIN-VC$_{\alpha,\beta}$. We will show that any approximation algorithm for MIN-VC$_{\alpha,\beta}$ also yields an approximation algorithm with almost the same approximation ratio for the problem of constructing a minimum size vertex cover for $\mathcal{G}_{\alpha,\beta}$ which contains the set $\Gamma$. This special version of the vertex cover problem for graphs $\mathcal{G}_{\alpha,\beta} = R_\beta(G_d)$ will be denoted as $\widehat{\text{MIN-VC}}_{\alpha,\beta}$.

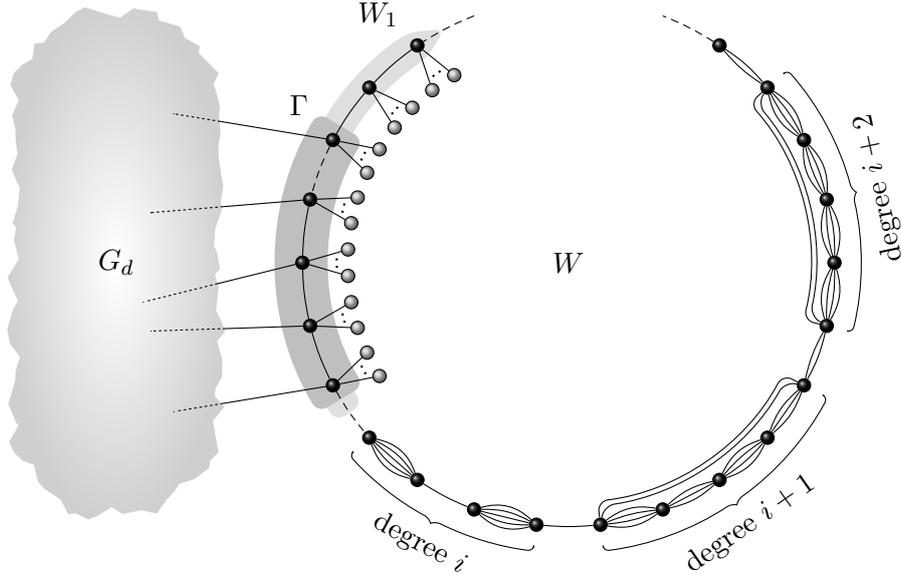

**Figure 5:** Embedding construction of the reduction $R_\beta$. Any $d$ bounded graph $G_d$ is attached to a multigraph wheel $W$ (more precisely, to a subset of vertices $\Gamma$) with a number of edges linear in $|G_d|$. The residual degrees of the power-law degree sequence are realized inside $W$ in a cyclic increasing order via degree 1 vertices and multi-edges. $W_1 \in V(W)$ denotes the subset of wheel nodes which have a degree 1 vertex attached.

In both cases, our polynomial time reduction from MIN-VC$_d$ to MIN-VC$_{\alpha,\beta}$ has the following general structure.

① Map $G_d$ to a multigraph $\mu(G_d)$.
*In the case $\beta > 1$, $\mu(G_d)$ is equal to $G_d$. In the case $\beta \leqslant 1$ and in the functional cases $\beta = 1 - \frac{1}{f(n)}$ and $\beta = 1 + \frac{1}{f(n)}$, we first apply our polynomial time reduction from MIN-VC$_d$ to MIN-VC$_{d+2}^{PM}$ - the Vertex Cover Problem restricted to $(d+2)$-bounded degree graphs which provide a perfect matching. This yields a graph $\widetilde{G}_{d+2}$ of size $4 \cdot |G_d|$ which contains a perfect matching $M$. Then we replace the edges of $M$ by multi-edges such as to increase the degree of nodes in $\widetilde{G}_{d+2}$ appropriately.*

② Choose the parameter $\alpha$ as small as possible such that $|\mu(G_d)| \leqslant |[x\Delta, y\Delta]|$
*(and possibly other constraints are satisfied as well)*

③ Construct the set of wheel-nodes $W$.
Assign to every node $v$ in $\mu(G_d) \cup W$ a node degree $\deg_{\alpha,\beta}(v)$ (the desired degree). Generate the multi-edges from $\mu(G_d)$ to the set $\Gamma \subset W$.
*The wheel will be constructed in such a way that wheel nodes of the same degree always form an induced connected subgraph in $W$.*

④ Connect the degree-1 nodes to the wheel $W$.

⑤ Construct edges inside $W$ such that the resulting multigraph is an $(\alpha, \beta)$-PLG.

In order to keep track of the node degrees and the edges being already constructed in steps ①-⑤ of this reduction, we keep track of the *residual degrees* $\deg_r(v)$ of nodes $v$ in the graph $G_{\alpha,\beta}$.

The completion step ⑤ of the reduction consists of the algorithm `Fill_Wheel` which we will describe now. This algorithm gets as an input the set of wheel nodes $W$ with residual degrees $\deg_r(w), w \in W$. It generates the missing edges degree-wise in a cyclic order. If $w_{j,1}, \ldots, w_{j,n_j}$ are the nodes of degree $\deg_{\alpha,\beta}(w_{j,l}) = j$ in the wheel $W$, then the following invariant will be maintained.

**Invariant 1.** *In every stage of the construction, for every* $j \in \{1, \ldots, \Delta\}, \deg_r(w_{j,1}) \leqslant \ldots \leqslant \deg_r(w_{j,n_j})$ *and* $\deg_r(w_{j,n_j}) - \deg_r(w_{j,1}) \leqslant 1$.

We are now ready to give the pseudo-code description of the algorithm `Fill_Wheel`.

---
**Algorithm 1:** `Fill_Wheel`

---
**Input**: The set of wheel nodes $\{w_{j,l}\} \in V(W)$ with $j \in \{3, \ldots, \Delta\}, l \in \{1, \ldots, n_j\}$ and residual degrees $\deg_r(w_{j,l})$.
**Output**: A graph $W$ with residual degrees $\deg_r(w_{j,l}) = 0$.

**for** $j = 3, \ldots, \Delta$ **do**
    **while** $\deg_r(w_{j,n_j}) > 0$ **do**
        choose $l$ min such that $\deg_r(w_{j,l})$ is max;
        **if** $l < n_j$ **then**
            generate edge $\{w_{j,l}, w_{j,l+1}\}$;
            $\deg_r(w_{j,l}) := \deg_r(w_{j,l}) - 1$;
            $\deg_r(w_{j,l+1}) := \deg_r(w_{j,l+1}) - 1$;
        **else if** $l = n_j, \deg_r(w_{j,1}) > 0$ **then**
            generate edge $\{w_{j,l}, w_{j,1}\}$;
            $\deg_r(w_{j,l}) := \deg_r(w_{j,l}) - 1$;
            $\deg_r(w_{j,1}) := \deg_r(w_{j,1}) - 1$;
        **else if** $l = n_j, \deg_r(w_{j,1}) = 0, j < \Delta$ **then**
            generate edge $\{w_{j,l}, w_{j+1,1}\}$;
            $\deg_r(w_{j,l}) := \deg_r(w_{j,l}) - 1$;
            $\deg_r(w_{j+1,1}) := \deg_r(w_{j+1,1}) - 1$;
        **else**
            take degree-1 node $w_1$ and generate edge $\{w_{j,l}, w_1\}$;
            $\deg_r(w_{j,l}) := \deg_r(w_{j,l}) - 1$;

---

## 4 Case $\beta > 1$

We will now consider the case when the parameter $\beta$ is in the range $1 < \beta < \beta_{\max} \approx 2.48$. We distinguish the subcases $1 < \beta < 2$, $\beta = 2$ and $2 < \beta < \beta_{\max}$ which differ by the choice of the intervals and the analysis of the construction.

## 4.1 Subcase $1 < \beta < 2$

Given a degree-$d$-bounded graph $G_d$, we construct $\mathcal{G}_{\alpha,\beta} = R_\beta(G_d)$ as follows: $W$ is the set of wheel nodes, $W_1 \subseteq W$ is the set of nodes $w \in W$ which are adjacent to at least one node of degree 1, and the set $\Gamma$ of neighbors of $G_d$ in the wheel satisfies $\Gamma \subseteq W_1 \subseteq W$. We want to choose $W_1 = [j_0, \Delta]$ as small as possible such as to meet the following requirements:

1. *Sufficient amount of node degree in the wheel:* $\lfloor e^\alpha \rfloor + n \leqslant \sum_{j=j_0}^{\Delta} (j-2) \cdot \left( \left\lfloor \frac{e^\alpha}{j^\beta} \right\rfloor - n_{j-1} \right)$

2. *Enough degree-$1$ nodes:* $\lfloor e^\alpha \rfloor \geqslant |\Gamma|$, which holds if $\lfloor e^\alpha \rfloor \geqslant \sum_{j=j_0}^{\Delta} \left( \left\lfloor \frac{e^\alpha}{j^\beta} \right\rfloor - n_{j-1} \right)$

3. *Node degrees of $G_d$:* $\left\lfloor \frac{e^\alpha}{j^\beta} \right\rfloor \geqslant n_{j-1} \quad (j = 2, \ldots, d+1)$

The first constraint ensures that a sufficient amount of node degree is available in the set $W_1$, such as to let all the nodes from $G_d$ and all the degree 1 nodes be adjacent to nodes from $W_1$. The second constraint guarantees that every node in the neighborhood $\Gamma$ of $G_d$ can be adjacent to at least one degree 1 node. Since we may assume that $G_d$ does not contain any node of degree 1 and since every node in $G_d$ will have one neighbor in $\Gamma$, the third constraint ensures that the degree distribution of the embedded graph $\mu(G_d)$ fits into the power-law distribution of the graph $\mathcal{G}_{\alpha,\beta}$.

**Lemma 2.** *If* $\left\lfloor \frac{e^\alpha}{(d+1)^\beta} \right\rfloor \geqslant n$, *then the constraint 3 is satisfied.*

Hence we choose $e^\alpha = (d+1)^\beta \cdot n$. In order to minimize the size of $W_1$, we want to choose $j_0$ as large as possible. This yields the requirement

1*. $\sum_{j=j_0}^{\Delta} (j-2) \cdot \left\lfloor \frac{e^\alpha}{j^\beta} \right\rfloor \geqslant n + \lfloor e^\alpha \rfloor$

which is equivalent to

$$e^\alpha \cdot \left( \left[ \frac{x}{2-\beta} \right]_{j_0}^{\Delta} - 2 \cdot \zeta(\beta) \right) \geqslant n + (d+1)^\beta \cdot n$$
$$\iff \quad \Delta^{2-\beta} - j_0^{2-\beta} \geqslant (2-\beta) \cdot \left( 1 + \frac{1}{(d+1)^\beta} + 2 \cdot \zeta(\beta) \right)$$

**Lemma 3.** *This inequality holds for $j_0 = \Delta - h(n)$ with $h(n) = \Delta^u$, $1 > u > \beta - 1$.*

*Proof.* (for the special case $\frac{1}{2-\beta} = l \in \mathbb{Z}$) In this case the requirement is equivalent to

$\Delta^{2-\beta} - \text{const.} \geqslant (\Delta - h(n))^{2-\beta} \Leftrightarrow \left( \Delta^{2-\beta} - \text{const.} \right)^{\frac{1}{2-\beta}} \geqslant \Delta - h(n) \Leftrightarrow \Delta - q\left(\Delta^{1/l}\right) \geqslant \Delta - h(n)$,

where $q$ is a polynomial of degree $l - 1 = \frac{1}{2-\beta} - 1 = \frac{\beta-1}{2-\beta}$. $\square$

Hence we can choose the parameter $u = \frac{1+(\beta-1)}{2} = \frac{\beta}{2}$ and $W_1 = [\Delta - \Delta^u, \Delta]$ and obtain $|W_1| = \sum_{j=\Delta-\Delta^u}^{\Delta} \left\lfloor \frac{e^\alpha}{j^\beta} \right\rfloor = o(|\mathcal{G}_{\alpha,\beta}|) = o(n)$. We are now ready to give the description of our reduction in the case $1 < \beta < 2$ in algorithm `Reduction`$_{\beta > 1}$.

**Resulting Lower Bound.** Suppose Min-VC$_d$ is hard to approximate within $1 + \varepsilon_d$. Suppose $\mathcal{A}_\beta$ is an approximation algorithm for Min-VC on $(\alpha, \beta)$-PLG with an approximation ratio $1 + \varepsilon_\beta$. This yields an approximation algorithm $\mathcal{A}_d$ for Min-VC$_d$. As we have seen before, we may assume

**Algorithm 2:** Reduction$_{\beta>1}$

**Input**: $G_d = (V, E)$ Degree-$d$-bounded graph with $V = \{v_1, \ldots, v_n\}$ such that
$2 \leqslant \deg_{G_d}(v_1) \leqslant \ldots \leqslant \deg_{G_d}(v_n) \leqslant d$

**Output**: $(\alpha, \beta)$-PLG $\mathcal{G}_{\alpha,\beta} = (V_{\alpha,\beta}, E_{\alpha,\beta})$ with $V_{\alpha,\beta} = V \cup W$

① **choose** $u = \frac{\beta}{2}$ ;

② **let** $\alpha = \min \left\{ \alpha' \,\middle|\, |[\Delta - \Delta^u, \Delta]| \geqslant n \text{ and } \left\lfloor \frac{e^\alpha}{(d+1)^\beta} \right\rfloor \geqslant n \right\}$ ;

③ *Generation of the Wheel Nodes:*

$n_j := \sharp$nodes of degree $j$ in $G_d$ $(j = 2, \ldots, d)$;

$V_{\alpha,\beta} := V_d \cup W$ $(W = \bigcup_{j=1}^{\Delta} W_{(j)}$ with $W_{(j)} = \left\{ w_{j,l} \,\middle|\, 1 \leqslant l \leqslant \left\lfloor \frac{e^\alpha}{j^\beta} \right\rfloor - n_{j-1} \right\}, j = 1, \ldots, \Delta)$;

④ *Rim Edges and Residual Degrees:*

Generate edges $\{w_{j,l}, w_{j,l+1}\}$ $(j = 2, \ldots, \Delta$ and $l = 1, \ldots, |W_{(j)}|))$;

Generate edges $\{w_{j,|W_{(j)}|}, w_{j+1,1}\}$ and one edge $\{w_{\Delta,1}, w_{2,1}\}$ $(j = 2, \ldots, \Delta)$;

**let** $\deg_r(w_{j,l}) := j - 2$ $(j = 2, \ldots, \Delta, 1 \leqslant l \leqslant |W_{(j)}|)$;

⑤ *Edges from $G_d$ to $W$:*

**for** $(c = 1, j = \Delta - \Delta^u; c \leqslant n; j++)$ **do**
    **for** $(l = 1; l < |W_{(j)}| \wedge c \leqslant n; l++, c++)$ **do**
        Generate an edge $\{v_c, w_{j,l}\}$ and set $\deg_r(w_{j,l}) := \deg_r(w_{j,l}) - 1$;

⑥ *Degree-1 Nodes:*

**for** $(j = \Delta - \Delta^u, c_1 = 0; j < \Delta \wedge c_1 < \lfloor e^\alpha \rfloor; j++)$ **do**
    **for** $(l = 1; l < |W_{(j)}| \wedge c_1 < \lfloor e^\alpha \rfloor; l = (l == |W_{(j)}| ? 1 : l+1))$ **do**
        Generate one edge $\{w_{1,c}, w_{j,l}\}$ and set $c_1 := c_1 + 1$, $\deg_r(w_{j,l}) := \deg_r(w_{j,l}) - 1$;

⑦ *Remaining Edges:*

Apply algorithm `Fill_Wheel`;

**let** $E_{\alpha,\beta}$ be the union of $E$ and the set of all edges generated in steps ④-⑦;

**return** $(V_{\alpha,\beta}, E_{\alpha,\beta})$

that on given input $R_\beta(G_d)$, algorithm $\mathcal{A}_\beta$ constructs a vertex cover $C_d \cup W_1 \cup \mathsf{OPT}(W \setminus W_1)$ and $\mathcal{A}_d$ on input $G_d$ returns the cover $C_d$. By assumption,

$$|C_d| + |W_1| + |\mathsf{OPT}(W \setminus W_1)| \leq (1 + \varepsilon_\beta) \cdot (|\mathsf{OPT}_d| + |W_1| + |\mathsf{OPT}(W \setminus W_1)|)$$
$$\Leftrightarrow \quad |C_d| \leq (1 + \varepsilon_\beta) \cdot |\mathsf{OPT}_d| + \varepsilon_\beta \cdot (|W_1| + |\mathsf{OPT}(W \setminus W_1)|)$$

Since $|W \setminus W_1| = \zeta(\beta) \cdot (d+1)^\beta \cdot n - n - e^\alpha - o(n)$, we obtain

$$|C_d| \leq (1 + \varepsilon_\beta) \cdot |\mathsf{OPT}_d| + \varepsilon_\beta \cdot \left( o(n) + \frac{(\zeta(\beta) - 1) \cdot (d+1)^\beta \cdot n - n - o(n)}{2} \right)$$
$$\leq (1 + \varepsilon_\beta) \cdot |\mathsf{OPT}_d| + \varepsilon_\beta \cdot \left( o(n) + \frac{(\zeta(\beta) - 1) \cdot (d+1)^\beta - 1 - o(1)}{2} \cdot d \cdot |\mathsf{OPT}_d| \right)$$
$$= |\mathsf{OPT}_d| \cdot \left( 1 + \varepsilon_\beta \cdot \left( 1 + o(1) + \frac{(\zeta(\beta) - 1) \cdot (d+1)^\beta - 1 - o(1)}{2} \cdot d \right) \right),$$

and hence the following result.

**Theorem 4.** *Suppose* MIN-VC$_d$ *is hard to approximate within approximation ratio* $1 + \varepsilon_d$. *Then for* $1 < \beta < 2$, MIN-VC$_{\alpha,\beta}$ *is hard to approximate within* $1 + \varepsilon_\beta = 1 + \frac{\varepsilon_d}{1 + d \cdot \frac{(\zeta(\beta) - 1) \cdot (d+1)^\beta - 1}{2}}$

### 4.2 Subcase $\beta = 2$

For this case we choose again $W_1 = [j_0, \Delta]$ and consider the corresponding optimization problem

*Minimize* $j_0$ *such that* $\left\lfloor \frac{e^\alpha}{(d+1)^\beta} \right\rfloor \geq n$ *and*

$$\sum_{j=j_0}^{\Delta} (j - 2) \cdot \left\lfloor \frac{e^\alpha}{j^\beta} \right\rfloor \geq n + \lfloor e^\alpha \rfloor \tag{1}$$

Suppose we first choose $\alpha$ such that $e^\alpha \geq (d+1)^\beta \cdot (n+1)$ (*) holds. Then the first constraint holds as well. For $\beta = 2$, assuming (*), we have the following chain of implications:

$$(1): \quad \sum_{j=j_0}^{\Delta} (j-2) \cdot \left\lfloor \frac{e^\alpha}{j^\beta} \right\rfloor \geq n + \lfloor e^\alpha \rfloor$$
$$\Leftarrow \quad \sum_{j=j_0}^{\Delta} \frac{e^\alpha}{j} - (\Delta - j_0 + 1) - 2 \sum_{j=j_0}^{\Delta} \frac{e^\alpha}{j^2} \geq n + e^\alpha$$
$$\Leftarrow \quad \sum_{j=j_0}^{\Delta} \frac{1}{j} - \frac{\Delta - j_0 + 1}{e^\alpha} - 2 \sum_{j=j_0}^{\Delta} \frac{1}{j^2} \geq \frac{1}{(d+1)^\beta} - \frac{1}{e^\alpha} + 1$$
$$\Leftarrow \quad \ln(\Delta) - \ln(j_0) - \frac{\Delta - j_0 + 1}{e^\alpha} - 2\zeta(2) \geq \frac{1}{(d+1)^\beta} - \frac{1}{e^\alpha} + 1$$

Hence we choose $j_0 = e^{\frac{\alpha}{2} - c}$ with $c = 2 + \frac{1}{(d+1)^\beta} + 2\zeta(2) = O(1)$. This implies $|W_1| = o(n)$, and thus we obtain the same lower bound as for the case $1 < \beta < 2$ as stated in [Theorem 4](#).

## 4.3 Subcase $\beta > 2$

For the case $2 < \beta < \beta_{\max} = \inf\{x|\zeta(x-1) - 2\zeta(x) \leq 0\}$ we consider the following construction. We construct the wheel in such a way that $W_1$ consists of all the wheel nodes of degree $\geq 3$. This yields $|W_1| = \sum_{j=3}^{\Delta} \left(\frac{e^\alpha}{j^\beta} - n_{j-1}\right)$ and $|W \setminus W_1| = \frac{e^\alpha}{2^\beta} - n_1 = \frac{e^\alpha}{2^\beta}$. We obtain

$$|C_d| \leq (1+\varepsilon_\beta)\mathsf{OPT}_d + \varepsilon_\beta \cdot \left(\frac{1}{2} \cdot \frac{e^\alpha}{2^\beta} + \zeta(\beta)e^\alpha - n - e^\alpha - \frac{e^\alpha}{2^\beta}\right)$$

$$= (1+\varepsilon_\beta)\mathsf{OPT}_d + \varepsilon_\beta \cdot e^\alpha \cdot \left(\frac{1}{2^{\beta-1}} + \zeta(\beta) - \frac{1}{(d+1)^\beta} - 1 - \frac{1}{2^\beta}\right)$$

$$\leq \mathsf{OPT}_d \cdot \left(1 + \varepsilon_\beta \cdot \left(1 + \left(\frac{1}{2^{\beta+1}} + \zeta(\beta) - 1 - \frac{1}{2^\beta} - \frac{1}{(d+1)^\beta}\right) \cdot (d+1)^\beta \cdot d\right)\right)$$

If instead we choose $|W_1| = \sum_{j=4}^{\Delta} \left(\frac{e^\alpha}{j^\beta} - n_{j-1}\right)$, then we obtain $|W \setminus W_1| = \frac{e^\alpha}{2^\beta} - \underbrace{n_1}_{=0} + \frac{e^\alpha}{3^\beta} - n_2$.

This yields

$$|C_d| \leq (1+\varepsilon_\beta)\mathsf{OPT}_d + \varepsilon_d \left(e^\alpha \left(\zeta(\beta) - 1 - \frac{1}{2^\beta} - \frac{1}{3^\beta}\right) + e^\alpha \cdot \frac{1 + 2^{-\beta} + 3^{-\beta}}{2} - n + \frac{n_1 + n_2}{2}\right)$$

$$= \mathsf{OPT}_d \cdot \left(1 + \varepsilon_\beta \cdot \left(1 + (d+1)^\beta \cdot d \cdot \left(\zeta(\beta) - \frac{1}{2} - \frac{1}{2^{\beta+1}} - \frac{1}{3^\beta \cdot 2}\right) - n + \frac{n_1 + n_2}{2}\right)\right).$$

We obtain the following theorem.

**Theorem 5.** *Suppose* MIN-VC$_d$ *is NP-hard to approximate within approximation ratio* $1 + \epsilon_d$. *Then, for* $2 < \beta < \beta_{\max} = \inf\{x|\zeta(x-1) - 2\zeta(x) \leq 0\} \approx 2.48$, MIN-VC$_{\alpha,\beta}$ *is hard to approximate within approximation ratio* $1 + \frac{\varepsilon_d}{1 + d(d+1)^\beta \left(\frac{1}{2^{\beta+1}} + \zeta(\beta) - 1 - \frac{1}{2^\beta} - \frac{1}{(d+1)^\beta}\right)}$.

## 5 Case $\beta \leq 1$

We consider now the case $0 < \beta \leq 1$. Again we construct a polynomial time reduction which embeds any $d$-bounded graph $G_d$ into an $(\alpha, \beta)$-PLG $\mathcal{G}_{\alpha,\beta}$. Since in the case $0 < \beta \leq 1$, the nodes of $G_d$ need to have high degree in $\mathcal{G}_{\alpha,\beta}$, we will first map $G_d$ to a $(d+2)$-bounded degree graph $\widetilde{G}_{d+2}$ which provides a perfect matching $M$. Then the edges of $M$ are duplicated in order to increase the degree of vertices in $\widetilde{G}_{d+2}$.

### 5.1 Min-VC in Bounded Degree Graphs which provide a Perfect Matching

We will now describe the polynomial time reduction from MIN-VC$_d$ to MIN-VC$_{d+2}^{PM}$. Given a graph $G_d$ of maximum degree $d$ with a vertex set $V = \{v_1, \ldots, v_n\}$, we construct the graph $\widetilde{G}_{d+2} = (\widetilde{V}, \widetilde{E})$ as follows:

① The set of vertices $\widetilde{V}$ consists of four disjoint copies of the vertex set $V$, namely $\widetilde{V} := V_1 \cup V_2 \cup V_3 \cup V_4$ with $V_i = \{v_{i,j}, 1 \leq j \leq n\}, i = 1, \ldots, 4$

② $\widetilde{E} := E_1 \cup E_2 \cup P$, where $G_d^1 = (V_1, E_1)$ and $G_d^2 = (V_2, E_2)$ are disjoint copies of $G_d$, i.e.

$$E_i = \{\{v_{i,j}, v_{i,l}\} | \{v_j, v_l\} \in E\}$$

③ $P := \bigcup_{j=1}^{n} \{\{v_{1,j}, v_{3,j}\}, \{v_{3,j}, v_{4,j}\}, \{v_{4,j}, v_{2,j}\}, \{v_{3,j}, v_{2,j}\}, \{v_{1,j}, v_{4,j}\}\}$

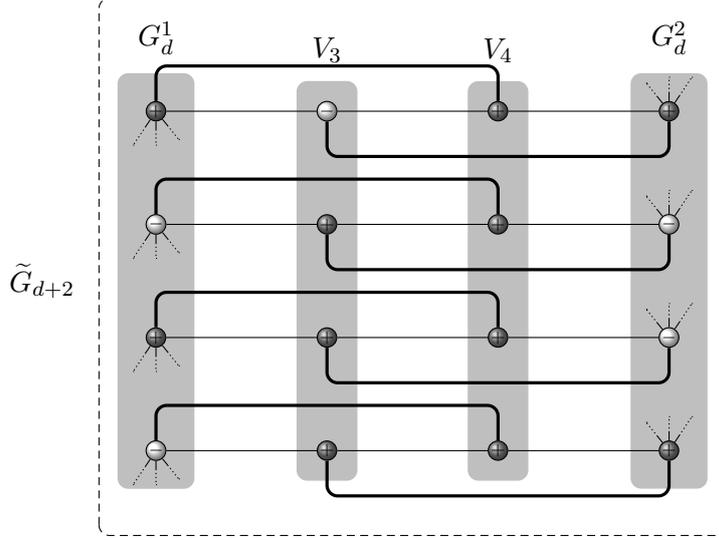

**Figure 6:** Example $\widetilde{G}_{d+2}$ after the construction step of reduction $R_{PM}$ that converts any $d$-bounded graph $G_d$ into a $(d+2)$-bounded graph which provides a perfect matching (e.g. via the **thick** edges of the set $P$). The nodes ⊕ denote covering vertices and ⊖ denote non-covering vertices.

The construction is shown in Figure 6. Let $R_{\mathsf{PM}}$ be this reduction, i.e. for any $d$-bounded degree graph $G_d$, $R_{\mathsf{PM}}(G) = \widetilde{G}_{d+2}$. Now suppose $C$ is a vertex cover in $\widetilde{G}_{d+2}$ with $C = C_1 \cup C_2 \cup C_3 \cup C_4$, where $C_i = C \cap V_i, 1 \leqslant i \leqslant 4$. We observe that $C_1$ and $C_2$ are vertex covers of $G^1$ and $G^2$, respectively. Furthermore, for every $j \in \{1, \ldots, n\}$ the following holds (see also Figure 6):

- If $v_{1,j} \in C$ and $v_{2,j} \in C$, then $C$ also contains one of the nodes $v_{3,j}, v_{4,j}$.
- If $v_{1,j} \notin C$ and $v_{2,j} \notin C$, then $C$ contains both nodes $v_{3,j}, v_{4,j}$.
- If $v_{1,j} \in C$ and $v_{2,j} \notin C$ – or vice versa –, then $C$ contains both nodes $v_{3,j}, v_{4,j}$.

Hence,

$$|C| = |C_1| + |C_2| + |C_1 \cap C_2| + 2 \cdot |V \setminus (C_1 \cup C_2)| + 2 \cdot |C_1 \triangle C_2|$$
$$= 3 \cdot |C_1 \triangle C_2| + 3 \cdot |C_1 \cap C_2| + 2 \cdot |V \setminus (C_1 \cup C_2)|$$
$$= 1 \cdot |C_1 \triangle C_2| + |C_1 \cap C_2| + 2 \cdot |V| \quad = \quad |C_1 \cup C_2| + 2 \cdot |V|.$$

This shows that a minimum is obtained by choosing $C_1 = C_2$ and minimizing the cardinality of this set. Hence we can restrict ourselves to vertex covers $C$ with the property $C_1 = C_2$ (formally: $C_1$ and $C_2$ being copies of the same vertex cover $C_d$ in $G_d$). Thus we obtain the following lemma.

**Lemma 6.** *There is a polynomial time algorithm $\mathcal{T}$ which transforms any vertex cover $C = C_1 \cup C_2 \cup C_3 \cup C_4$ of a graph $\widetilde{G}_{d+2} = R_{\mathsf{PM}}(G_d)$ into a vertex cover $C' = \mathcal{T}(G_d, C)$ of $\widetilde{G}_{d+2}$ such that $C'_1 = C'_2 = \mathrm{argmin}\{|C_1|, |C_2|\}$ and $|C'| = \min\{|C_1|, |C_2|\} + 2 \cdot |V| \leqslant |C|$.*

**Resulting lower bound.** Now suppose $\textsc{Min-VC}_d$ is hard to approximate within ratio $1 + \varepsilon_d$. Suppose $\mathcal{A}$ is a polynomial time $(1+\varepsilon)$-approximation algorithm for $\textsc{Min-VC}_{d+2}^{PM}$. Then the algorithm $\mathcal{B}$ which on input $G_d$ constructs the vertex cover $\widetilde{C} = \mathcal{T} \circ \mathcal{A} \circ R_{\mathsf{PM}}(G_d) = \bigcup_{i=1}^{4} \widetilde{C}_i$ and then returns $\widetilde{C}_1$ is an approximation algorithm for $\textsc{Min-VC}_d$. We have,

$$|\widetilde{C}| \leqslant (1+\varepsilon) \cdot \mathsf{OPT}(\widetilde{G}_{d+2}) = (1+\varepsilon) \cdot (\mathsf{OPT}(G_d) + 2n). \tag{2}$$

Furthermore due to the proof of the previous Lemma 6, $|\widetilde{C}| = |\widetilde{C}_1| + 2n$. Thus from Equation 2 we obtain

$$|\widetilde{C}_1| \leq (1+\varepsilon) \cdot \mathsf{OPT}(G_d) + \varepsilon \cdot 2n \leq ((1+\varepsilon) + d \cdot 2\varepsilon) \cdot \mathsf{OPT}(G_d),$$

where the second inequality holds due to the fact that $\mathsf{OPT}(G_d) \geq \frac{n}{d}$. Thus it must be that $\varepsilon_d \leq (d \cdot 2 + 1) \cdot \varepsilon$. Finally we note that if $G_d$ has maximum degree at most $d$, then the maximum degree of $\widetilde{G}_{d+2} = R_{\mathsf{PM}}(G_d)$ is at most $d+2$. Altogether we obtain the following result.

**Theorem 7.** *If* $\textsc{Min-VC}_d$ *is hard to approximate within approximation ratio* $1 + \varepsilon_d$, *then* $\textsc{Min-VC}_{d+2}^{PM}$ *is hard to approximate within approximation ratio* $1 + \frac{\varepsilon_d}{1+2d}$.

## 5.2 Subcase $0 < \beta < 1$

We consider now the subcase $0 < \beta < 1$. We start by giving an estimate of the cardinality of node intervals $[x\Delta, y\Delta]$ in $(\alpha, \beta)$-PLG. Although the rounding errors in the case $\beta < 1$ can be of order $\Theta(|\mathcal{G}_{\alpha,\beta}|)$, our estimates will enable us to choose the interval sizes appropriately and to obtain explicit lower bounds for $\textsc{Min-VC}_{\alpha,\beta}$.

**Lemma 8.** *(Sizes of Intervals) Let* $0 < \beta < 1$ *and let* $\mathcal{G}_{\alpha,\beta} = (V, E)$ *be an* $(\alpha, \beta)$-*PLG. For every* $0 < x < y < 1$, *the cardinality of the interval* $[x\Delta, y\Delta] = \{v \in V | x\Delta \leq \deg_{\alpha,\beta}(v) \leq y\Delta\}$ *is in*

$$\left[\frac{\Delta}{1-\beta}\left(y^{1-\beta} - x^{1-\beta}\right) - (y-x)\Delta - 1, \frac{\Delta}{1-\beta}\left(y^{1-\beta} - x^{1-\beta}\right) + \left(\frac{1}{x^\beta} - \frac{1}{y^\beta}\right)\right].$$

*Proof.* We first observe that

$$|[x\Delta, y\Delta]| = \sum_{j=x\Delta}^{y\Delta} \left\lfloor \frac{e^\alpha}{j^\beta} \right\rfloor \in \left[\sum_{j=x\Delta}^{y\Delta} \frac{e^\alpha}{j^\beta} - (y-x)\Delta - 1, \sum_{j=x\Delta}^{y\Delta} \frac{e^\alpha}{j^\beta}\right]$$

Since, for $0 < \beta < 1$, $\frac{1}{\chi^\beta}$ is a convex function, we obtain

$$\sum_{j=x\Delta}^{y\Delta} \frac{e^\alpha}{j^\beta} \in \left[e^\alpha \cdot \int_{x\Delta}^{y\Delta} \chi^{-\beta} d\chi, \ e^\alpha \left(\int_{x\Delta}^{y\Delta} \chi^{-\beta} d\chi + \left(\frac{1}{(x\Delta)^\beta} - \frac{1}{(y\Delta)^\beta}\right)\right)\right]$$

$$= \left[e^\alpha \cdot \left[\frac{\chi^{1-\beta}}{1-\beta}\right]_{x\Delta}^{y\Delta}, \ e^\alpha \cdot \left(\left[\frac{\chi^{1-\beta}}{1-\beta}\right]_{x\Delta}^{y\Delta} + \left(\frac{1}{(x\Delta)^\beta} - \frac{1}{(y\Delta)^\beta}\right)\right)\right]$$

$$= \left[\frac{\Delta}{1-\beta}\left(y^{1-\beta} - x^{1-\beta}\right), \frac{\Delta}{1-\beta}\left(y^{1-\beta} - x^{1-\beta}\right) + \left(\frac{1}{x^\beta} - \frac{1}{y^\beta}\right)\right]$$

□

We want to choose $0 \leq x < y < z \leq 1$ in such a way that the vertices of $\widetilde{G}_{d+2}$ will be contained in the interval $[x\Delta, y\Delta]$ and $\Gamma$ in the interval $(y\Delta, z\Delta]$. The preceding Lemma 8 shows that, in order to achieve $|\mathcal{G}_{\alpha,\beta}| = O(n)$, we have to choose $y = \Omega(1)$. The next lemma shows that we can even choose $y = 1 - o(1)$ and $x = o(1)$, which then implies $|[x\Delta, y\Delta]| = (1-o(1))|\mathcal{G}_{\alpha,\beta}|$.

**Lemma 9.** *Let* $x = \frac{d+1}{\Delta}, y = \left(1 + \frac{1}{\Delta^{1-\beta}}\right)^{-\frac{1}{2-\beta}}$ *and* $z = 1$. *Then* $|[x\Delta, y\Delta]| = (1-o(1))|\mathcal{G}_{\alpha,\beta}|$. *Furthermore,*

$$\sum_{j=y\Delta+1}^{\Delta} \left\lfloor \frac{e^\alpha}{j^\beta} \right\rfloor \cdot (j-2) \geq (d-1) \cdot n \tag{3}$$

*Proof.* Due to the previous Lemma 8,

$$|[x\Delta, y\Delta]| \geq \frac{\Delta}{1-\beta}\left(\left(1+\frac{1}{\Delta^{1-\beta}}\right)^{-\frac{1-\beta}{2-\beta}} - \left(\frac{d+1}{\Delta}\right)^{1-\beta}\right) - y\Delta$$

Hence we cannot apply Lemma 8 directly to the interval $[x\Delta, y\Delta]$, since the rounding error $(y-x)\Delta$ is of order $\Omega(\Delta) = \Omega(|\mathcal{G}_{\alpha,\beta}|)$. Instead, we apply the lemma to the complement of $[x\Delta, y\Delta]$ in $\mathcal{G}_{\alpha,\beta}$:

$$|[x\Delta, y\Delta] = |\mathcal{G}_{\alpha,\beta}| - |(y\Delta, \Delta]| - \sum_{j=1}^{d}\left\lfloor\frac{e^\alpha}{j^\beta}\right\rfloor$$

$$\geq |\mathcal{G}_{\alpha,\beta}| - \frac{\Delta}{1-\beta}\underbrace{\left(1 - \left(1+\frac{1}{\Delta^{1-\beta}}\right)^{-\frac{1-\beta}{2-\beta}}\right)}_{=o(1)} - \left(1+\frac{1}{\Delta^{1-\beta}}\right)^{\frac{\beta}{2-\beta}} + 1 - \Theta(e^\alpha)$$

Since $|\mathcal{G}_{\alpha,\beta}| = \Theta(\Delta)$, we obtain $|[x\Delta, y\Delta] = (1-o(1)) \cdot |\mathcal{G}_{\alpha,\beta}|$. Now we show that for this choice of $x$ and $y$, the main inequality (3) of this lemma holds as well.

$$\sum_{j=y\Delta+1}^{\Delta}\left\lfloor\frac{e^\alpha}{j^\beta}\right\rfloor \cdot (j-2) \geq \sum_{j=y\Delta+1}^{\Delta}\frac{e^\alpha}{j^\beta}(j-2) - \underbrace{(\Delta - y\Delta - 1)}_{=(1-y)\Delta-1=o(\Delta)}$$

$$= \sum_{j=y\Delta}^{\Delta}\frac{e^\alpha}{j^\beta}(j-2) - \frac{e^\alpha}{(y\Delta)^\beta} \cdot (y\Delta - 2) - o(\Delta)$$

$$\geq e^\alpha \cdot \left(\int_{y\Delta}^{\Delta}\chi^{1-\beta}d\chi - 2\cdot\int_{y\Delta}^{\Delta}\chi^{-\beta}d\chi - 2\left(\frac{1}{y^\beta}-1\right)\right) - \frac{e^\alpha}{(y\Delta)^\beta} \cdot (y\Delta - 2) - o(\Delta)$$

(now using the fact that $2(y^{-\beta} - 1)e^\alpha = o(\Delta)$)

$$= e^\alpha \cdot \left(\left[\frac{\chi^{2-\beta}}{2-\beta}\right]_{y\Delta}^{\Delta} - 2\cdot\left[\frac{\chi^{1-\beta}}{1-\beta}\right]_{y\Delta}^{\Delta}\right) - \frac{e^\alpha}{(y\Delta)^\beta} \cdot (y\Delta - 2) - o(\Delta)$$

$$= e^\alpha \cdot \underbrace{\left(\frac{\Delta^{2-\beta}}{2-\beta} - \frac{(y\Delta)^{2-\beta}}{2-\beta} - \frac{2\Delta^{1-\beta}}{1-\beta} + \frac{2(y\Delta)^{1-\beta}}{1-\beta}\right)}_{1=z=\left(1+\frac{1}{\Delta^{1-\beta}}\right)^{\frac{1}{2-\beta}}y} - \underbrace{\frac{e^\alpha}{(y\Delta)^\beta} \cdot (y\Delta - 2)}_{=\Theta(y^{1-\beta}\Delta)} - o(\Delta)$$

$$= e^\alpha \cdot \left(\frac{\left(1+\frac{1}{\Delta^{1-\beta}}-1\right) \cdot y^{2-\beta} \cdot \Delta^{2-\beta}}{2-\beta} - \Theta\left(\Delta^{1-\beta}\right)\right) - \Theta(\Delta)$$

(observe that $e^\alpha \cdot \Delta^{1-\beta} = e^\alpha \cdot e^{\alpha \cdot \frac{1-\beta}{\beta}} = \Delta$)

$$= e^\alpha \cdot \frac{y^{2-\beta} \cdot \Delta^{1+\beta}}{2-\beta} - \Theta(\Delta) = e^\alpha \cdot \frac{y^{2-\beta} \cdot \Delta^{1+\beta}}{2-\beta} \cdot (1-o(1))$$

Therefore, $\sum_{j=y\Delta+1}^{\Delta}\left\lfloor\frac{e^\alpha}{j^\beta}\right\rfloor \cdot (j-2) = \omega(\Delta)$. Since $(d-1)n = O(n) = O(\Delta)$, inequality (3) holds. $\square$

The next lemma shows that $|\mathcal{G}_{\alpha,\beta}| = (1+o(1)) \cdot n$.

**Lemma 10.** *Let $\alpha = \min\{\alpha'| \, |[x\Delta, y\Delta]| \geqslant n\}$, where $x = \frac{d+1}{\Delta}$ and $y = \left(1 + \frac{1}{\Delta^{1-\beta}}\right)^{-\frac{1}{2-\beta}}$. Then, $|[x\Delta, y\Delta]| = (1 + o(1)) \cdot n$.*

*Proof.* For $\alpha = \min\{\alpha'| \, |[x\Delta, y\Delta]| \geqslant n\}$,

$$|[x\Delta, y\Delta]| = (1 - \tau(\alpha)) \cdot \frac{\Delta}{1-\beta} \geqslant n,$$

where $\tau(\alpha) = o(1)$. Hence

$$\alpha = \min\left\{\alpha' \,\middle|\, \alpha' \geqslant \beta \cdot \left(\ln\left(\frac{1-\beta}{1-\tau(\alpha')}\right) + \ln(n)\right) \text{ and } \frac{\Delta}{1-\beta} \in \mathbb{Z}\right\}$$
$$= \beta \cdot \left(\ln\left(\frac{1-\beta}{1-o(1)}\right) + \ln(n)\right) + o(1),$$

which implies $|\mathcal{G}_{\alpha,\beta}| = \frac{\Delta}{1-\beta} = n \cdot \frac{1-\beta}{1-o(1)} \cdot \frac{e^{o(1)}}{1-\beta} = (1 + o(1))n$ □

We obtain a polynomial time reduction from MIN-VC$_d$ to MIN-VC$_{\alpha,\beta}$ for the case $0 < \beta < 1$ in algorithm `Reduction`$_{\beta \leqslant 1}$.

**Resulting Lower Bound.** Suppose MIN-VC$_{d+2}^{PM}$ is NP-hard to approximate within approximation ratio $1 + \varepsilon_0$. Let $G_d = (V_d, E_d)$ be the $d$-bounded degree graph and $\mathcal{G}_{\alpha,\beta} = R(G_d)$. Let $|V_d| = n$. Let $W$ denote the wheel in $\mathcal{G}_{\alpha,\beta}$. We have $|W| \leqslant c \cdot n$. Let $\Gamma$ be the neighborhood of $G_d$ in $\mathcal{G}_{\alpha,\beta}$ with $|\Gamma| = \gamma = o(n)$.

Suppose $\mathcal{A}$ is a polynomial time approximation algorithm for MIN-VC in power-law graphs with an approximation ratio $1 + \varepsilon$. On input $\mathcal{G}_{\alpha,\beta} = R(G_d)$, algorithm $\mathcal{A}$ constructs a vertex cover $C = C_d \cup C_W$ with $C_d = C \cap V_d$ and $C_W = C \cap W$. Since no vertex from $W$ can cover any edge in $G_d$, $C_d$ is a vertex cover of $G_d$.

Let OPT denote a minimum cost vertex cover of $\mathcal{G}_{\alpha,\beta}$. Let $\mathsf{OPT}_d = \mathsf{OPT} \cap V_d$ and $\mathsf{OPT}_W = \mathsf{OPT} \cap W$. Then $|C| \leqslant (1+\varepsilon) \cdot \mathsf{OPT}$, $|C_d| > \frac{n}{d}$ and

$$|\mathsf{OPT}| = |\mathsf{OPT}_d \cup \mathsf{OPT}_W| \leqslant |\mathsf{OPT}_d| + |\mathsf{OPT}_W| + |\Gamma|$$
$$\leqslant (1 + o(1)) \cdot (|\mathsf{OPT}_d \cup \mathsf{OPT}_W|) = (1 + o(1)) \cdot |\mathsf{OPT}|$$

Hence the approximation algorithm $\mathcal{B}$ which on input $\mathcal{G}_{\alpha,\beta}$ first computes the cover $C = \mathcal{A}(\mathcal{G}_{\alpha,\beta})$ and then replaces $C_W$ by the union of $\Gamma$ and an optimum vertex cover for $W \setminus \Gamma$ (which can be computed efficiently by dynamic programming) has approximation ratio $(1 + \varepsilon_0) \cdot (1 + o(1))$ for the instances $\mathcal{G}_{\alpha,\beta} = G_d \cup W = R(G_d)$ of MIN-VC.

We will now show that algorithm $\mathcal{B}$ has also a similar approximation ratio for a slightly modified optimization problem:

**Problem 1** ($\widehat{\text{MIN-VC}}$).

**Input:** $d$-bounded degree graph $G_d$.

**Output:** Vertex cover $C$ for $R(G_d) = G_d \cup W$ such that $\Gamma \subseteq C$.

**Objective:** Minimize $|C|$.

Let $\widehat{\mathsf{OPT}}(G_d)$ denote an optimum solution for instance $G_d$ of this modified optimization problem. Furthermore let $\mathsf{OPT}(G_d)$ and $\mathsf{OPT}(W \setminus \Gamma)$ denote minimum cost vertex covers for $G_d$ and the graph $W \setminus \Gamma$, respectively. Then

$$\widehat{\mathsf{OPT}}(G_d) = \mathsf{OPT}(G_d) \cup \Gamma \cup \mathsf{OPT}(W \setminus \Gamma).$$

**Algorithm 3:** $\text{Reduction}_{\beta \leqslant 1}$

**Input**: $G_d = (V, E)$ a $d$-bounded graph with $V = \{v_1, \ldots, v_n\}$ such that
$\quad 2 \leqslant \deg_{G_d}(v_1) \leqslant \ldots \leqslant \deg_{G_d}(v_n) \leqslant d$
**Output**: $(\alpha, \beta)$-PLG $\mathcal{G}_{\alpha,\beta} = (V_{\alpha,\beta}, E_{\alpha,\beta})$ with $V_{\alpha,\beta} = V \cup W$

① *Generate the Perfect Matching Graph:*
**let** $\widetilde{G}_{d+2} = R_{\mathsf{PM}}(G_d) = (\widetilde{V}, \widetilde{E})$;
**let** $\widetilde{M} = \{e_1, \ldots, e_{2n}\}$ be a perfect matching in $\widetilde{G}_{d+2}$;
② *Choose $\alpha, x$ and $y$:*
**let** $\alpha := \min \left\{ \alpha' \,\middle|\, |[x\Delta, y\Delta]| \geqslant \left|\widetilde{G}_{d+2}\right| \right\}$;
**let** $x := \frac{d+1}{\Delta}$ and $y := \left(1 + \frac{1}{\Delta^{1-\beta}}\right)^{-\frac{1}{2-\beta}}$;
③ *Duplicate Edges and generate Residual Degrees of $\widetilde{G}_{d+2}$:*
Assign degrees $\deg_{\alpha,\beta}(v_i)$ to the nodes $v_i$ of $\widetilde{G}_{d+2}$
such that $x\Delta \leqslant \deg_{\alpha,\beta}(v_i) \leqslant y\Delta$, respecting the $(\alpha, \beta)$ power-law;
**let** $\deg_r(v_i) = \deg_{\alpha,\beta}(v_i) - \deg_{\widetilde{G}_{d+2}}(v_i)$;
**for** $i = 1, \ldots, 2n$ **do**
$\quad$ **let** $e_i = \{v_{i_1}, v_{i_2}\}$;
$\quad$ Replace $e_i$ by $\min\{\deg_r(v_{i_1}), \deg_r(v_{i_2})\} - 1$ parallel edges;
$\quad$ Update $\deg_r(v_{i_1}), \deg_r(v_{i_2})$ accordingly;

**let** $\mu(G_d)$ the resulting multigraph;
④ *Generation of the Wheel Nodes:*
$n_j := \sharp \text{nodes } v_i \text{ with } \deg_{\alpha,\beta}(v_i) = j \text{ in } \widetilde{G}_{d+2} \quad (j = 2, \ldots, \Delta)$;
$V_{\alpha,\beta} := \widetilde{V}_{d+2} \cup W \quad (W = \bigcup_{j=1}^{\Delta} W_{(j)}$ with
$W_{(j)} = \left\{ w_{j,l} \,\middle|\, 1 \leqslant l \leqslant \left\lfloor \frac{e^\alpha}{j^\beta} \right\rfloor - n_{j-1} \right\} \quad (j = 1, \ldots, \Delta)$;
⑤ *Rim Edges and Residual Degrees:*
Generate edges $\{w_{j,l}, w_{j,l+1}\} \quad (j = 2, \ldots, \Delta \text{ and } l = 1, \ldots, |W_{(j)}|)$;
Generate edges $\{w_{j,|W_{(j)}|}, w_{j+1,1}\}$ and one edge $\{w_{\Delta,1}, w_{2,1}\} \quad (j = 2, \ldots, \Delta)$;
**let** $\deg_r(w_{j,l}) := j - 2 \quad (j = 2, \ldots, \Delta, \, 1 \leqslant l \leqslant |W_{(j)}|)$;
⑥ *Edges from $\mu(G_d)$ to $W$:*
**for** $(c = 1, j = y\Delta + 1; \, c \leqslant 4n; \, j\texttt{++})$ **do**
$\quad$ **for** $(l = 1; \, l < |W_{(j)}| \wedge c \leqslant 4n; \,)$ **do**
$\quad\quad$ Generate $\min\{\deg_r(v_c), \deg_r(w_{j,l})\}$ parallel edges between $v_c$ and $w_{j,l}$;
$\quad\quad$ Update $\deg_r(v_c), \deg_r(w_{j,l})$ accordingly;
$\quad\quad$ **if** $\deg_r(w_{j,l}) = 0$ **then** $j\texttt{++}$;
$\quad\quad$ **if** $\deg_r(v_c) = 0$ **then** $c\texttt{++}$;

⑦ *Degree-1 Nodes:*
Connect the nodes $w_{1,l}$ to $\bigcup_{y\Delta < j \leqslant \Delta} W_{(j)}$; $\quad$ /* c.f. step ⑥ of $\texttt{Reduction}_{\beta>1}$ */
⑧ *Remaining Edges:*
Apply algorithm `Fill_Wheel`;
**let** $E_{\alpha,\beta}$ be the set of edges generated in steps ③-⑧;
**return** $(V_{\alpha,\beta}, E_{\alpha,\beta})$;

We observe that if $\mathsf{OPT} = \mathsf{OPT}_d \cup \mathsf{OPT}_W$ is an optimum vertex cover for $\mathcal{G}_{\alpha,\beta} = R(G_d)$, then

$$|\mathsf{OPT}_d| \leqslant |\mathsf{OPT}(G_d)| + |\Gamma|$$

We have $|\widehat{\mathsf{OPT}}(G_d)| \leqslant |\mathsf{OPT}| + |\Gamma| + O(1) = (1 + o(1)) \cdot |\mathsf{OPT}|$. We show that algorithm $\mathcal{B}'$ has approximation ratio $(1 + o(1)) \cdot (1 + \varepsilon)$ for the modified optimization problem. Then we can conclude:

$$|C_d \cup \Gamma \cup \mathsf{OPT}(W \setminus \Gamma)| \leqslant (1 + o(1)) \cdot (1 + \varepsilon) \cdot \widehat{\mathsf{OPT}}(G_d)$$
$$= (1 + o(1)) \cdot (1 + \varepsilon) \cdot |\mathsf{OPT}(G_d) \cup \Gamma \cup \mathsf{OPT}(W \setminus \Gamma)|$$

which yields

$$|C_d| \leqslant (1 + o(1)) \cdot (1 + \varepsilon) \cdot |\mathsf{OPT}(G_d)| + \varepsilon \cdot |\Gamma \cup \mathsf{OPT}(W \setminus \Gamma)|$$

Now since $|\Gamma \cup \mathsf{OPT}(W \setminus \Gamma)| \leqslant |W| \leqslant c \cdot n$ and $|\mathsf{OPT}(G_d)| \geqslant \frac{n}{d}$, we obtain $|\Gamma \cup \mathsf{OPT}(W \setminus \Gamma)| \leqslant c \cdot d \cdot |\mathsf{OPT}(G_d)|$ and therefore

$$|C_d| \leqslant (1 + o(1)) \cdot (1 + \varepsilon + \varepsilon \cdot c \cdot d) \cdot |\mathsf{OPT}(G_d)|$$

and thus $\varepsilon \cdot (1 + c \cdot d) \geqslant (1 - o(1)) \cdot \varepsilon_0$. For our choice of the parameters $x, y, z$ we obtain $c = o(1)$, and thus the following theorem holds.

**Theorem 11.** *If* MIN-VC$_d$ *is hard to approximate within ratio* $1 + \epsilon_d$, *then for* $0 < \beta < 1$, MIN-VC$_{\alpha,\beta}$ *is hard to approximate within approximation ratio* $1 + \frac{\epsilon_d}{1+2d}$.

## 5.3 Subcase $\beta = 1$

The case $\beta = 1$ differs from the case $0 < \beta < 1$ by how we choose the intervals $[x\Delta, y\Delta]$ and $(y\Delta, z\Delta]$. Nevertheless, we will obtain the same lower bound as in the case $\beta = 1$.

**Lemma 12.** (Sizes of Intervals)
*Let* $\mathcal{G}_{\alpha,\beta}$ *be an* $(\alpha,\beta)$-*PLG with* $\beta = 1$. *Then for all* $0 < x < y \leqslant 1$, *the size of the interval* $[x\Delta, y\Delta] = \{v \in V(\mathcal{G}_{\alpha,\beta}) | x\Delta \leqslant \mathsf{deg}(v) \leqslant y\Delta\}$ *satisfies*

$$|[x\Delta, y\Delta]| \in \left[(\ln(y) - \ln(x) - (y - x + 1)) \cdot e^\alpha, \ e^\alpha \cdot (\ln(y) - \ln(x)) + \left(\frac{1}{x} - \frac{1}{y}\right)\right]$$

*Proof.* First we give a bound for the rounding error:

$$\sum_{j=x\Delta}^{y\Delta} \frac{e^\alpha}{j} - (y - x + 1) \cdot \Delta \leqslant \sum_{j=x\Delta}^{y\Delta} \left\lfloor \frac{e^\alpha}{j} \right\rfloor \leqslant \sum_{j=x\Delta}^{y\Delta} \frac{e^\alpha}{j}$$

For the term $\sum_{j=x\Delta}^{y\Delta} j^{-1}$ we get the following bounds:

$$\sum_{j=x\Delta}^{y\Delta} \frac{1}{j} \in \left[\int_{x\Delta}^{y\Delta} \chi^{-1} d\chi, \int_{x\Delta}^{y\Delta} \chi^{-1} d\chi + \left(\frac{1}{x\Delta} - \frac{1}{y\Delta}\right)\right]$$
$$= \left[\ln(y\Delta) - \ln(x\Delta), \ \ln(y\Delta) - \ln(x\Delta) + \left(\frac{1}{x\Delta} - \frac{1}{y\Delta}\right)\right]$$
$$= \left[\ln(y) - \ln(x), \ \ln(y) - \ln(x) + \left(\frac{1}{x\Delta} - \frac{1}{y\Delta}\right)\right]$$

Thus the lemma follows. □

In the case $\beta < 1$ we have mapped $G_d$ to a subinterval $[x\Delta, y\Delta) = \{v \in V(\mathcal{G}_{\alpha,\beta}) | x\Delta \leqslant \deg(v) < y\Delta\}$, where $0 < x < y < 1$ and $x, y$ are constant. However, in the case $\beta = 1$ the size of such an interval is $\Theta(e^\alpha)$ which is $o(|\mathcal{G}_{\alpha,1}|)$. This means we have to choose the interval bounds in a different way.

**Lemma 13.** *Let $\mathcal{G}_{\alpha,\beta} = (V, E)$ be an $(\alpha, 1)$-PLG. For $0 \leqslant c' < c \leqslant 1$ and $x = e^{-(1-c')\alpha}$, $y = e^{-(1-c)\alpha}$ with $c, c'$ being constant, the size of the set $[x\Delta, y\Delta] = \{v \in V | x\Delta \leqslant \deg(v) \leqslant y\Delta\}$ satisfies*
$$|[x\Delta, y\Delta]| = (1 - o(1))(c - c') \cdot \alpha e^\alpha = \Theta(\alpha \cdot e^\alpha).$$

*Proof.* Due to the preceding lemma,
$$|[x\Delta, y\Delta]| \geqslant e^\alpha \cdot (\ln(y) - \ln(x) - (y - x + 1))$$
$$= e^\alpha \cdot ((1 - c') - (1 - c))\alpha - O(1)) = (1 - o(1)) \cdot (c - c')\alpha e^\alpha$$
□

The next lemma shows that if we choose $\alpha$ as small as possible such as to be able to embed $G_d$ into the interval $[x\Delta, y\Delta]$, then we obtain $|[x\Delta, y\Delta]| = (1 + o(1))n$.

**Lemma 14.** *For $x = \frac{1}{e^{(1-c')\cdot\alpha}}, y = \frac{1}{e^{(1-c)\cdot\alpha}}$ and $\alpha = \min\{\alpha' | |[x\Delta, y\Delta]| \geqslant n\}$,*
$$n \leqslant |[x\Delta, y\Delta]| \leqslant n + t(n)$$
*where $t(n) = \lfloor y\Delta \rfloor - \lceil x\Delta \rceil + O(1)$, especially $t(n) = o(n)$.*

*Proof.* The equation $t(n) = \lfloor y\Delta \rfloor - \lceil x\Delta \rceil + O(1)$ follows directly from the choice of $x$ and $y$. It remains to show that $t(n) = o(n)$. From $t(n) = o(|\mathcal{G}_{\alpha,\beta}|)$ and $|[x\Delta, y\Delta]| = \Theta(|\mathcal{G}_{\alpha,\beta}|)$ we obtain $t(n) = o(|[x\Delta, y\Delta]|)$. The inequality $n \leqslant |[x\Delta, y\Delta]| \leqslant n + t(n)$ then implies $n = \Theta(|[x\Delta, y\Delta]|)$, whence $t(n) = o(n)$. □

Finally we show that we can choose $\Gamma = (y\Delta, z\Delta]$ with $z = c''\Delta$, $c'' = c + o(1)$.

**Lemma 15.** *Let $0 \leqslant c' < c < 1$ be constants and $c'' = c + \frac{1}{\alpha}$, and let $x = e^{(c'-1)\alpha}$, $y = e^{(c-1)\alpha}$ and $z = e^{(c''-1)\alpha}$. Then $|(y\Delta, z\Delta]| = o(\alpha e^\alpha)$ and*
$$\sum_{j=y\Delta+1}^{z\Delta} \left\lfloor \frac{e^\alpha}{j} \right\rfloor \cdot (j - 2) = \omega(|\mathcal{G}_{\alpha,\beta}|)$$

*Proof.* Using the preceding lemma,
$$|(y\Delta, z\Delta]| = |[y\Delta, z\Delta]| - \left\lfloor \frac{e^\alpha}{y\Delta} \right\rfloor$$
$$\leqslant e^\alpha \cdot (c'' - c) \cdot \alpha + e^{(1-c)\alpha} - e^{(1-c'')\alpha} = e^\alpha + e^{(1-c)\alpha} - e^{(1-c'')\alpha} = o(|\mathcal{G}_{\alpha,\beta}|)$$

Furthermore,
$$\sum_{j=y\Delta+1}^{z\Delta} \left\lfloor \frac{e^\alpha}{j} \right\rfloor \cdot (j - 2) \geqslant \sum_{j=y\Delta+1}^{z\Delta} \frac{e^\alpha}{j} \cdot (j - 2) - (z - y)\Delta \cdot (z\Delta - 2)$$
$$= \sum_{j=y\Delta}^{z\Delta} \frac{e^\alpha}{j} \cdot (j - 2) - (z - y)\Delta \cdot (z\Delta - 2) - \frac{e^\alpha}{y\Delta}$$
$$\geqslant e^\alpha \cdot ((z - y + 1)\Delta - 2(\ln(z\Delta) - \ln(y\Delta)))$$
$$- (z - y)\Delta \cdot (z\Delta - 2) - \frac{e^\alpha}{y\Delta} - 2\left(\frac{1}{y\Delta} - \frac{1}{z\Delta}\right)$$
$$= (1 - o(1)) \cdot e^{2\alpha} = \omega(\alpha \cdot e^\alpha)$$
□

Thus for $\beta = 1$, we obtain basically the same polynomial time reduction from MIN-VC$_d$ to MIN-VC$_{\alpha,1}$ as in the case $0 < \beta < 1$. The only difference is how we choose the parameters $x, y, z$. We obtain the following result.

**Theorem 16.** *If* MIN-VC$_d$ *is hard to approximate within ratio* $1+\epsilon_d$, *then for* $\beta = 1$, MIN-VC$_{\alpha,\beta}$ *is hard to approximate within approximation ratio* $1 + \frac{\epsilon_d}{1+2d}$.

# 6 Functional Case $\beta = 1 - \frac{1}{f(n)}$

We consider now the case of $\beta = 1 - \frac{1}{f(n)}$ being a function of the number of nodes $n$, converging to 1 from below. Here is a precise description of the model.

**Definition 2.** *($(\alpha, \beta)$-PLG for $\beta = 1 - \frac{1}{f(n)}$)*
Let $f(n)$ be a monotone increasing unbounded function. For $\beta = 1 - \frac{1}{f(n)}$, an $(\alpha, \beta)$-PLG $\mathcal{G}_{\alpha,\beta}$ is a multigraph with $n$ nodes which has the following properties:

(1) The maximum degree of $\mathcal{G}_{\alpha,\beta}$ is $\Delta_f = \lfloor e^{\alpha/\beta} \rfloor$.

(2) There are $\lfloor \frac{e^\alpha}{j^\beta} \rfloor$ nodes of degree $j$ ($j = 1, \ldots, \Delta_f$)

Thus an $(\alpha, \beta)$-PLG $\mathcal{G}_{\alpha,\beta}$ satisfies the equation $n = \sum_{j=1}^{\Delta} \lfloor \frac{e^\alpha}{j^\beta} \rfloor$. In this section we will show that the approximation lower bound for MIN-VC$_{\alpha,1}$ also holds for the functional case $\beta = 1 - \frac{1}{f(n)}$. We achieve this by showing that the crucial parameters (maximum degree, sizes of intervals) of $(\alpha, \beta)$-PLG in the functional case converge to those in the case $\beta = 1$. Let us start by giving an outline of the main steps. We may first ask when do the single terms $\frac{1}{i^{1-f(n)-1}}$ converge to $\frac{1}{i}$. The differences (Local Error) are

$$\frac{1}{i^{\frac{f(n)-1}{f(n)}}} - \frac{1}{i} = \frac{i^{\frac{1}{f(n)}} - 1}{i} \leqslant \frac{n^{\frac{1}{f(n)}} - 1}{i}$$

We have $\log\left(n^{1/f(n)}\right) = \frac{\log(n)}{f(n)}$, hence for $f(n) = \omega(\log n)$ the nominator converges to 0. Another estimate of the local error can be obtained as follows: $\frac{i^{1/f(n)} - 1}{i} \leqslant \frac{\Delta^{1/f(n)} - 1}{i} = \left(\frac{n}{f(n)}\right)^{\frac{1}{f(n)}} - 1$. Then we are going to deal with the global error, i.e. we consider the sum

$$\sum_{i=1}^{\Delta} \frac{e^\alpha}{i^\beta} = e^\alpha \cdot \sum_{i=1}^{e^{\alpha \cdot \frac{f(n)}{f(n)-1}}} \frac{1}{i^{\frac{f(n)-1}{f(n)}}} \stackrel{!}{=} n$$

We will show that this sum will differ from the according sum of the terms $\frac{1}{i}$ by an amount of $\alpha \cdot e^\alpha \cdot \frac{f(n)}{f(n)-1} \cdot \left(e^{\frac{\alpha}{f(n)-1}} - 1\right)$, which is a lower oder term provided $f(n) = \omega(\alpha)$. We will also give bound on the rounding error when we replace the terms $\lfloor e^\alpha / j^\beta \rfloor$ by their fractional counterparts.

**Detailed Description.** We let $\Delta_f = \left\lfloor e^{\alpha \cdot \frac{f(n)}{f(n)-1}} \right\rfloor$ and $\Delta = \lfloor e^\alpha \rfloor$. Thus $\Delta_f = \left\lfloor e^\alpha \cdot e^{\frac{\alpha}{f(n)-1}} \right\rfloor =$
$(1 + o(1)) \cdot \Delta$ provided $f$ satisfies $f(n) = \omega(\alpha)$. In the case $\beta = 1 - \frac{1}{f(n)}$ we have $n = \sum_{j=1}^{\Delta_f} \lfloor \frac{e^\alpha}{j^\beta} \rfloor$.
We want to give upper and lower bounds for this term. Since $\Delta_f = (1 + o(1))\Delta$, we obtain

$$\sum_{j=1}^{\Delta} \frac{e^\alpha}{j^\beta} - (1+o(1))\Delta \leqslant \sum_{j=1}^{\Delta_f} \frac{e^\alpha}{j^\beta} - \Delta_f \leqslant \sum_{j=1}^{\Delta_f} \left\lfloor \frac{e^\alpha}{j^\beta} \right\rfloor \leqslant \sum_{j=1}^{\Delta_f} \frac{e^\alpha}{j^\beta}$$

The right-hand side of this inequality (the upper bound) can be further bounded as follows:

$$\sum_{j=1}^{\Delta_f} \frac{e^\alpha}{j^\beta} = \sum_{j=1}^{\Delta} \frac{e^\alpha}{j^\beta} + (\Delta_f - \Delta) \cdot \frac{e^\alpha}{\Delta^\beta} = \sum_{j=1}^{\Delta} \frac{e^\alpha}{j^\beta} + o(1) \cdot \Delta \cdot \frac{e^\alpha}{\Delta^{1-\frac{1}{f(n)}}}$$

$$= \sum_{j=1}^{\Delta} \frac{e^\alpha}{j^\beta} + o(1) \cdot e^\alpha \cdot \Delta^{\frac{1}{f(n)}} = \sum_{j=1}^{\Delta} \frac{e^\alpha}{j^\beta} + o(1) \cdot e^\alpha \cdot e^{\frac{\alpha}{f(n)}}$$

$$= \sum_{j=1}^{\Delta} \frac{e^\alpha}{j^\beta} + o(1) \cdot (1 + o(1)) \cdot e^\alpha$$

where the last equality holds again due to the fact that $f(n) = \omega(\alpha)$. Thus we obtain the following lemma.

**Lemma 17.** *If $f(n) = \omega(\alpha)$, then*

$$\sum_{j=1}^{\Delta} \frac{e^\alpha}{j^\beta} - (1 + o(1))\Delta \leqslant \sum_{j=1}^{\Delta_f} \left\lfloor \frac{e^\alpha}{j^\beta} \right\rfloor = n \leqslant \sum_{j=1}^{\Delta} \frac{e^\alpha}{j^\beta} + o(1) \cdot \Delta$$

We also need similar bounds for the sizes of intervals. It will turn out that we can restrict ourselves to intervals $[a(n), b(n)]$ in $(\alpha, \beta)$-PLG where $b(n) \leqslant \lfloor e^\alpha \rfloor$ (instead of $b(n) \leqslant \Delta_f = \left\lfloor e^{\alpha \cdot \frac{f(n)}{f(n)-1}} \right\rfloor$).

**Lemma 18.** *Suppose $f(n) = \omega(\alpha)$. Let $a, b \colon \mathbb{N} \to \mathbb{N}$ such that for all $n$, $1 \leqslant a(n) < b(n) \leqslant \lfloor e^\alpha \rfloor$. Then*

$$\sum_{j=a(n)}^{b(n)} \frac{e^\alpha}{j^\beta} - (b(n) - a(n) + 1) \leqslant |[a(n), b(n)]| = \sum_{j=a(n)}^{b(n)} \left\lfloor \frac{e^\alpha}{j^\beta} \right\rfloor \leqslant \sum_{j=a(n)}^{b(n)} \frac{e^\alpha}{j^\beta}$$

The next lemma gives the desired bounds for the sizes of intervals in the functional case $\beta = 1 - \frac{1}{f(n)}$. The upper and lower bounds are sums of terms $\frac{e^\alpha}{j}$ instead of $\frac{e^\alpha}{j^{\beta_f}}$. Afterwards we will use this result to show that we can actually choose the same parameters $x\Delta, y\Delta, z\Delta$ as in the case $\beta = 1$.

**Lemma 19.** *(Convergence of Sizes of Intervals)*
*For each pair of functions $a, b \colon \mathbb{N} \to \mathbb{N}$ with $1 \leqslant a(n) < b(n) \leqslant \Delta_f$,*

$$\sum_{j=a(n)}^{b(n)} \frac{e^\alpha}{j} - \frac{n}{\log(n)} \leqslant |[a(n), b(n)]| \leqslant (1 + \varepsilon(n)) \cdot \sum_{j=a(n)}^{b(n)} \frac{e^\alpha}{j} \tag{4}$$

*where $\varepsilon(n) = n^{\frac{1}{f(n)-1}}$. Especially $\varepsilon(n) = o(1)$ for $f(n) = \omega(\log(\alpha))$, which implies*

$$(1 - o(1)) \sum_{j=a(n)}^{b(n)} \frac{e^\alpha}{j} \leqslant |[a(n), b(n)]| \leqslant (1 + o(1)) \sum_{j=a(n)}^{b(n)} \frac{e^\alpha}{j}$$

*Proof.* The second inequality in Equation 4 holds if for each $j$, $\frac{1}{j^{1-\frac{1}{f(n)}}} \leqslant (1 + \varepsilon(n)) \cdot \frac{1}{j}$, i.e. $j^{\frac{1}{f(n)}} \leqslant 1 + \varepsilon(n) \Leftrightarrow j \leqslant (1 + \varepsilon(n))^{f(n)} \Leftarrow \Delta_f \leqslant (1 + \varepsilon(n))^{f(n)}$. This last inequality is equivalent to $e^{\frac{\alpha}{f(n)-1}} \leqslant 1 + \varepsilon(n)$ and (taking logarithm) $\frac{\alpha}{f(n)-1} \leqslant f(n) \cdot \ln(1 + \varepsilon(n))$. Since $f(n) = \omega(\alpha)$, we obtain $\varepsilon(n) \geqslant e^{\frac{\alpha}{f(n)-1}} - 1 = n^{\frac{1-o(1)}{f(n)-1}}$, hence for $\varepsilon(n) = n^{\frac{1}{f(n)-1}}$ we obtain

$$\sum_{j=a(n)}^{b(n)} \frac{e^\alpha}{j} - (b(n) - a(n) + 1) \leqslant |[a(n), b(n)]| \leqslant (1 + \varepsilon(n)) \cdot \sum_{j=a(n)}^{b(n)} \frac{e^\alpha}{j}$$

We have $b(n) - a(n) + 1 \leqslant \Delta_f = (1 + o(1)) \cdot \Delta$. Now we consider the last inequality for the special case when $a(n) = 1$ and $b(n) = \Delta_f$. We obtain:

$$(1 + o(1)) \sum_{j=1}^{\Delta} \frac{e^\alpha}{j} - (1 + o(1))\Delta = \sum_{j=1}^{\Delta_f} \frac{e^\alpha}{j} - (1 + o(1))\Delta \leqslant n = |[1, \Delta_f]|,$$

which implies $n = (1 + o(1)) \cdot \alpha \cdot e^\alpha$. □

Now let $x = \frac{1}{e^{(1-c')\alpha}}$, $y = \frac{1}{e^{(1-c)\alpha}}$, $z = \frac{1}{e^{(1-c'')\alpha}}$ with $0 \leqslant c' < c < c'' = c + \frac{1}{\alpha}$. Combining Lemma 19 with the proof of Lemma 12, we obtain

$$(c - c')\alpha e^\alpha - \frac{n}{\log(n)} \leqslant |[x\Delta, y\Delta]| \qquad \leqslant \left((c - c')\alpha e^\alpha + \frac{1}{x\Delta} - \frac{1}{y\Delta}\right) \cdot (1 + o(1))$$

$$e^\alpha - \frac{n}{\log(n)} - \frac{e^\alpha}{(y\Delta)^{\beta_f}} \leqslant |(y\Delta, z\Delta]| \qquad \leqslant \left(e^\alpha + \frac{1}{z\Delta} - \frac{1}{y\Delta}\right) \cdot (1 + o(1))$$

which yields $|[x\Delta, y\Delta]| = (1 \pm o(1))(c - c')\alpha e^\alpha$ and $|(y\Delta, z\Delta]| = o(\alpha e^\alpha)$. Now choose $c = 1 - \frac{1}{\alpha}$ and $c' = \frac{d+1}{\Delta}$. Then

$$\sum_{j=y\Delta+1}^{\Delta} \left\lfloor \frac{e^\alpha}{j^{\beta_f}} \right\rfloor \cdot (j-2) \geqslant \sum_{j=y\Delta}^{\Delta} \frac{e^\alpha}{j^{\beta_f - 1}} - \frac{1-y}{2} \cdot \Delta^2 - 2 \cdot (1 + o(1))e^\alpha$$

$$\geqslant e^\alpha \cdot \int_{y\Delta}^{z\Delta} \chi^{1/f(n)} d\chi - \left(1 + o(1) - \frac{1}{e}\right) \frac{\Delta^2}{2}$$

$$= e^\alpha \cdot \frac{f(n)}{f(n)+1} \cdot \left(\Delta^{1+\frac{1}{f(n)}} - (y\Delta)^{1+\frac{1}{f(n)}}\right) - \left(1 + o(1) - \frac{1}{e}\right) \frac{\Delta^2}{2}$$

$$= \frac{1 - e^{-1} - o(1)}{2} \cdot e^{\alpha \cdot \left(2+\frac{1}{f(n)}\right)} \quad = \quad \omega\left(|\mathcal{G}_{\alpha,\beta_f}|\right)$$

Hence we can map $G'_d$ to the interval $[x\Delta, y\Delta]$ and choose $\Gamma \subseteq (y\Delta, z\Delta]$ (note that we choose $\Delta = \lfloor e^\alpha \rfloor$ instead of $\Delta_f = \lfloor e^{\alpha/\beta_f} \rfloor$) and obtain the same hardness result as in the case $\beta \leqslant 1$ when $\beta$ is a constant. We have the following result.

**Theorem 20.** *Suppose* MIN-VC$_d$ *is hard to approximate within ratio* $1 + \epsilon_d$. *Let* $f\colon \mathbb{N} \to \mathbb{N}$ *be a function such that* $f(n) = \omega(\log(n))$. *Then for* $\beta_f = 1 - \frac{1}{f(n)}$, *the problem* MIN-VC$_{\alpha,\beta_f}$ *is hard to approximate within approximation ratio* $1 + \frac{\epsilon_d}{1+2d}$.

# 7 Functional Case $\beta = 1 + \frac{1}{f(n)}$

It turns out that even in the functional case $\beta = 1 + \frac{1}{f(n)}$, we obtain the same hardness result as in the case $\beta \leqslant 1$. This is especially interesting since we have the phase transition at $\beta = 1$, which is also reflected by our hardness results for $\beta$ being a constant. Again, our result is based on an estimate of the sizes of intervals $[a(n), b(n)]$.

**Lemma 21.** *Let $\beta_f = 1 + \frac{1}{f(n)}$ with $f(n) = \omega(\alpha)$, and let $\Delta_f = \left\lfloor e^{\alpha/\beta_f} \right\rfloor$. Then for each $j \in \{1, \ldots, \Delta_f\}$,*
$$\frac{1}{j^{\beta_f}} \in \left[\frac{1}{j} - \tau(n), \frac{1}{j}\right],$$
*where $\tau(n) = \frac{2^{\frac{1}{f(n)}}-1}{2^{1+\frac{1}{f(n)}}}$. Especially $\tau(n) \longrightarrow 0$ as $n$ goes to infinity.*

*Proof.* We have $\frac{1}{j^{\beta_f}} \in \left[\frac{1}{j} - t(j), \frac{1}{j}\right]$ for $t(j) = \frac{1}{j} - \frac{1}{j^\beta} = \frac{j^{\beta_f-1}-1}{j^{\beta_f}}$. It suffices to show that $t(j) \leqslant \tau(n) = \frac{2^{1/f(n)}-1}{2^{1+1/f(n)}}$. Since for fixed $n$, the derivative of the function $x \mapsto \frac{x^{1/f(n)}-1}{x^{1+1/f(n)}} = x^{-1} - x^{-1-\frac{1}{f(n)}}$ is equal to $-x^{-2} + \left(1 + \frac{1}{f(n)}\right) x^{-2-\frac{1}{f(n)}} \leqslant 0$, the inequality $t(j) \leqslant t(2) = \tau(n)$ holds, and thus the lemma follows. □

If we combine this result with our techniques from previous sections, the resulting estimate for sizes of intervals is rather weak. For the number of nodes of the $\alpha, \beta$-PLG, we obtain
$$\frac{\alpha}{1 + \frac{1}{f(n)}} e^\alpha - \frac{e^{(1+\frac{f(n)}{f(n)+1})\alpha + O(1)}}{2^{1+1/f(n)}} \leqslant |[1, \Delta_f]| \leqslant (1 \pm o(1))\alpha e^\alpha$$

We will make use of the following estimate of the local rounding errors.

**Lemma 22.** *For every $j \in \{1, \ldots, \Delta_f\}$, $\frac{1}{j^{1+\frac{1}{f(n)}}} \in \left[\frac{1}{n^{\frac{1}{f(n)}}} \cdot \frac{1}{j}, \frac{1}{j}\right]$.*

*Proof.* We just observe that $\frac{1}{j^{1+\frac{1}{f(n)}}} = \frac{1}{j} \cdot \frac{1}{j^{1/f(n)}}$, and the function $x \mapsto \frac{1}{x^{1/f(n)}}$ is monotone decreasing. □

This gives the following estimate of sizes of intervals.

**Lemma 23.** *(Sizes of Intervals)*
*In the case $\beta_f = 1 + \frac{1}{f(n)}$, for any $1 \leqslant a < b \leqslant \Delta_f = \left\lfloor e^{\alpha/\beta_f} \right\rfloor$, the size of the interval $[a, b] = \{v \in V(\mathcal{G}_{\alpha,\beta})|\; a \leqslant \deg(v) \leqslant b\}$ is in*
$$\left[\frac{1}{n^{\frac{1}{f(n)}}} \cdot e^\alpha \cdot (\ln(b) - \ln(a)) - (b - a + 1),\; e^\alpha \cdot (\ln(b) - \ln(a)) + e^\alpha \cdot \left(\frac{1}{a} - \frac{1}{b}\right)\right]$$

Since for $f(n) = \omega(\ln(n))$, we have convergence $n^{\frac{1}{f(n)}} \longrightarrow 1 \;(n \to \infty)$, we obtain the following estimates.

**Corollary 1.** *For $f(n) = \omega(\ln(n))$, the number of nodes of an $(\alpha, \beta_f)$-PLG $\mathcal{G}_{\alpha,\beta_f}$ satisfies $|[1, \Delta_f]| = (1 \pm o(1)) \cdot \alpha \cdot e^\alpha$. Furthermore, for the parameters $x = \frac{1}{e^{(1-c')\alpha}}, y = \frac{1}{e^{(1-c)\alpha}}, z = \frac{1}{e^{(1-c'')\alpha}}$ with $0 \leqslant c' < c < c'' = c + \frac{1}{\alpha}$, $|[x\Delta_f, y\Delta_f]| = (1 \pm o(1)) \cdot |[1, \Delta_f]|$ and $|(y\Delta_f, z\Delta_f]| = (1 \pm o(1))e^\alpha = o(|\mathcal{G}_{\alpha,\beta_f}|)$.*

Now we show that if we choose the parameters $x, y, z$ such that $z = 1$, then the amount of node-degree in the interval $(y\Delta_f, z\Delta_f]$ suffices to connect all the nodes from $\widetilde{G}_d$ as well as the

degree-1 nodes. Namely, choose $c'' = 1, c = 1 - \frac{1}{\alpha}$. Then we obtain

$$\sum_{j=y\Delta_f+1}^{\Delta_f} \left\lfloor \frac{e^\alpha}{j^{\beta_f}} \right\rfloor (j-2) \geqslant \sum_{j=y\Delta_f}^{\Delta_f} \frac{e^\alpha}{j} \cdot \frac{1}{n^{1/f(n)}} \cdot (j-2) \ - (j-2)$$

$$= \frac{e^\alpha}{n^{1/f(n)}} \cdot (1-y)\Delta_f \ - \frac{2e^\alpha}{n^{1/f(n)}} \sum_{j=y\Delta_f+1}^{\Delta_f} \frac{1}{j}$$

$$= (1-o(1)) \cdot \left(1 - \frac{1}{e}\right) \cdot e^{\alpha\left(1+\frac{f(n)}{1+f(n)}\right)} \quad = \omega\left(\left|\mathcal{G}_{\alpha,\beta_f}\right|\right)$$

Finally we obtain the following result.

**Theorem 24.** *Suppose* Min-VC$_d$ *is hard to approximate within ratio* $1 + \epsilon_d$. *Let* $f \colon \mathbb{N} \to \mathbb{N}$ *be a function such that* $f(n) = \omega(\log(n))$. *Then for* $\beta_f = 1 + \frac{1}{f(n)}$, *the problem* Min-VC$_{\alpha,\beta_f}$ *is hard to approximate within approximation ratio* $1 + \frac{\epsilon_d}{1+2d}$.

## 8 Further Research

In this paper we have given explicit lower bounds for the approximability of Min-VC in connected $(\alpha, \beta)$-PLG. It remains an important open question to close the gaps between inapproximability and approximability bounds of the underlying problems. We also believe that our results for the two functional cases $\beta = 1 \pm \frac{1}{f(n)}$ can be extended to hold for any $\beta_f = \beta \pm \frac{1}{f(n)}$ with $0 < \beta < \beta_{\max} \approx 2.48$. It would also be interesting to study the approximation complexity of various network design problems on Power-Law Graphs, e.g. the *Steiner Tree Problem* and related problems.